\documentclass[a4paper,12pt]{article}
\pdfoutput=1
\usepackage{amsmath}
\usepackage{amsfonts}
\usepackage{amssymb}
\usepackage{graphicx, rotating}
\usepackage{epsfig}
\usepackage{latexsym}
\usepackage{graphicx}
\usepackage{color}
\usepackage{amsmath,bm,amssymb}
\usepackage[normalem]{ulem}

\usepackage{color}
\usepackage[english]{babel}
\usepackage{hyperref}
\usepackage[numbers,sort&compress]{natbib}

\newcommand{\Slash}[1]{{\ooalign{\hfil#1\hfil\crcr\raise.167ex\hbox{/}}}}

\newcommand{\beq}{\begin{equation}}  \newcommand{\eeq}{\end{equation}}
\newcommand{\bef}{\begin{figure}}  \newcommand{\eef}{\end{figure}}
\newcommand{\bec}{\begin{center}}  \newcommand{\eec}{\end{center}}
  
\newcommand{\laq}[1]{\label{eq:#1}}  

\newcommand{\Eq}[1]{Eq.~(\ref{eq:#1})}
\newcommand{\Eqs}[1]{Eqs.~(\ref{eq:#1})}
\newcommand{\eq}[1]{(\ref{eq:#1})}
\newcommand{\Sec}[1]{Sec.~\ref{chap:#1}}

\newcommand{\vev}[1]{ \left\langle {#1} \right\rangle }

\newcommand{\lac}[1]{\label{chap:#1}}
\newcommand{\SU}[1]{{\rm SU{#1} } }

\newcommand{\pdvo}[1]{\frac{\partial}{\partial #1} }
\newcommand{\pdvt}[2]{\frac{\partial #1}{\partial #2} }
\newcommand{\pqty}[1]{\left( #1 \right)}

\newcommand{\sqty}[1]{\left[ #1 \right]}
\def\({\left(}
\def\){\right)}

\def\O{\mathcal{O}}

\newcommand{\AND}{~{\rm and}~}
\newcommand{\EV}{ {\rm \, eV} }

\newcommand{\MEV}{ {\rm \, MeV} }
\newcommand{\GEV}{ {\rm \, GeV} }
\newcommand{\TEV}{ {\rm \, TeV} }

\def\a{\alpha}
\def\b{\beta}

\def\d{\delta}
\def\e{\epsilon}
\def\f{\phi}
\def\g{\gamma}
\def\h{\theta}
\def\k{\kappa}
\def\l{\lambda}

\def\r{\rho}
\def\s{\sigma}

\def\D{\Delta}
\def\G{\Gamma}
\def\H{\Theta}
\def\L{\Lambda}

\def\P{\Psi}

\def\tl{\tilde}
\def\*{\dagger}

\def\fpa{$\delta \phi_{\parallel}$ }
\def\fpe{$\delta \phi_{\perp}$ }

\begin{document}

\begin{center}

\hfill   TU-1207\\

\vspace{1.5cm}

{\Large\bf  QCD Axion Hybrid Inflation}
\vspace{1.5cm}

{\bf Yuma Narita, Fuminobu Takahashi, and Wen Yin}

\vspace{12pt}
\vspace{1.5cm}
{\em 
Department of Physics, Tohoku University,  
Sendai, Miyagi 980-8578, Japan 
\vspace{5pt}
}

\vspace{1.5cm}
\abstract{
When the inflaton is coupled to the gluon Chern-Simons term for successful reheating, mixing between the inflaton and the QCD axion is generally expected given the solution of the strong CP problem by the QCD axion. This is particularly natural if the inflaton is a different, heavier axion. 
We propose a scenario in which the QCD axion plays the role of the inflaton by mixing with heavy axions. In particular, if the energy scale of inflation is lower than the QCD scale, a hybrid inflation is realized where the QCD axion plays the role of the inflaton in early stages. We perform detailed numerical calculations to take account of the mixing effects. Interestingly, the initial misalignment angle of the QCD axion, which is usually a free parameter, is determined by the inflaton dynamics. It is found to be close to $\pi$ in simple models. This is the realization of the pi-shift inflation proposed in previous literature, and it shows that QCD axion dark matter and inflation can be closely related. {The heavy axion {may} be probed by future accelerator experiments.}}

\end{center}
\clearpage

\setcounter{page}{1}
\setcounter{footnote}{0}

\section{Introduction}

Inflation not only solves initial value problems such as the horizon  and the flatness problems inherent in Big Bang cosmology but also naturally generates primordial density fluctuations that become the seeds of the structure of the universe~\cite{Starobinsky:1980te,Guth:1980zm,Sato:1980yn,Linde:1981mu,Albrecht:1982wi}. Observations of the Cosmic Microwave Background (CMB) and large-scale structure strongly suggest that inflation occurred in the early universe. 

In the inflationary universe, inflation and reheating are equally important. It is necessary to convert the energy of the inflaton that caused inflation into a hot plasma including Standard Model (SM) particles and connect it to hot Big Bang cosmology. In a simple scenario, the inflaton has a direct coupling with the SM particles. For example, suppose that the inflaton $\phi$ interacts with the Chern-Simons term of gluons as $ \phi G \tilde{G}$.
This is a natural possibility  if the inflaton $\phi$ is an axion. In fact, many axions are considered to appear in the low energy from string theory, which is known as the axiverse or axion landscape, and various cosmological roles of axions are discussed in the literature~\cite{Witten:1984dg, Svrcek:2006yi,Conlon:2006tq,Arvanitaki:2009fg,Acharya:2010zx, Higaki:2011me, Cicoli:2012sz, Higaki:2014pja,Higaki:2014mwa,Daido:2016tsj,Stott:2017hvl,Bachlechner:2018gew,Demirtas:2018akl,Nakagawa:2020eeg,Reig:2021ipa,Demirtas:2021gsq} (see
Refs.\,\cite{Jaeckel:2010ni,Ringwald:2012hr,Arias:2012az,Graham:2015ouw,Marsh:2015xka,Irastorza:2018dyq,DiLuzio:2020wdo} for reviews). In particular, since the potential of the axion is protected from perturbative radiative corrections, it is possible to realize a flat potential necessary for slow-roll inflation, and many axion inflation models have been proposed~\cite{Freese:1990rb,Adams:1992bn, Czerny:2014wza, Czerny:2014xja,Croon:2014dma}.

Among many axions, the QCD axion is predicted in the Peccei-Quinn (PQ) mechanism, which is one of the plausible solutions to the strong CP problem in SM. The key ingredient is that the QCD axion, {$a$}, is coupled to the Chern-Simons term of the gluons~\cite{Peccei:1977hh,Peccei:1977ur,Weinberg:1977ma,Wilczek:1977pj} as {$aG\tl G$,} and it acquires mass from non-perturbative effects of QCD. 
Thus, if the inflaton had a coupling with the gluon Chern-Simons term for successful reheating, the inflaton and the QCD axion would mix below the QCD scale. In particular, 
if the inflationary Hubble parameter $H_{\rm inf}$ is below the QCD scale, they already mix during inflation.
This can be understood by noting that the Gibbons-Hawking temperature during inflation is given by $H_{\rm inf}/(2\pi)$.
Such low-scale inflation can be realized by another axion with a potential that has a flat plateau near the top of the potential~\cite{Daido:2017wwb, Daido:2017tbr, Takahashi:2019qmh, Takahashi:2021tff, Takahashi:2023vhv}, which is a low-scale realization of multi-natural inflation~\cite{Czerny:2014wza, Czerny:2014xja, Czerny:2014qqa, Higaki:2014sja}.

The interplay between the QCD axion and the inflaton
 has been studied in the context of the production of the
QCD axion dark matter (DM)~\cite{Daido:2017wwb,Takahashi:2019pqf, Takahashi:2019qmh, Kobayashi:2019eyg, Nakagawa:2020eeg}. 
However, the inflaton dynamics was considered separately as a background, and the mixing due to non-perturbative effects of QCD during inflation was not taken into account. 
In this paper, we focus on scenarios where the QCD axion mixes with the inflaton through the coupling to the gluon Chern-Simons term, investigating whether the QCD axion plays a significant role in the dynamics of inflation.
Specifically, we explore low-scale inflation  driven by two axions, $a$ and $\phi$ both coupled to QCD, where $a$ is the QCD axion and $\phi$ is another heavier axion.  We will demonstrate that the inflaton dynamics resemble those considered in Refs.~\cite{Peloso:2015dsa,Daido:2017wwb}. The behavior is also somewhat analogous to hybrid inflation\cite{Copeland:1994vg,Dvali:1994ms,Linde:1997sj}, although in our case, slow-roll inflation could persist long after the inflaton trajectory changes direction. The most common inflation model for this low-energy scale is quartic hilltop inflation. However, the predicted spectral index is known to be lower than the value determined by the CMB observations.  We will show that the QCD axion drives inflation in the early stages, and the mixing between the QCD axion and the inflaton gradually changes the potential and the trajectory of the inflaton, which can better explain the CMB observations.

There are some interesting implications of the inflation driven by the QCD axion.
First, the quality problem of the PQ symmetry~\cite{Misner:1957mt,Banks:1988yz, Barr:1992qq,Kamionkowski:1992mf,Holman:1992us,Kallosh:1995hi} is now related to the condition for slow-roll inflation, since an explicit PQ symmetry breaking could spoil the slow-roll inflation. Another is that the initial misalignment angle of the QCD axion is fixed by the inflaton dynamics. We will see that, in simple models, it is set close to $\pi$, which enhances the abundance of QCD axion DM by the anharmonic effects~\cite{Lyth:1991ub,Kobayashi:2013nva}. This favors a relatively small decay constant of the QCD axion, $f_a \sim 10^9$\,GeV, to explain DM.  The other, heavier axion {has a slightly larger mass than the sensitivity limits of the currently planned accelerator experiment but may be explored in the future. We clarify the conditions for the experiment, which turns out to be feasible.}

Finally, let us comment on the difference with Ref.~\cite{Takahashi:2019pqf, Takahashi:2021tff}. 
The idea of the QCD hybrid inflation was first raised in Ref.\,\cite{Takahashi:2019pqf}, but it was argued that the scenario is in tension with the supernova bound on the QCD axion from a na\"{i}ve estimate by integrating out the heavier waterfall field $\f$ {than} $H_{\rm inf}.$ In this paper, we perform analytical and numerical analyses of the inflation dynamics by fully considering the two-field evolution and show that there is a wide range of viable parameters. {This scenario is also  different from \cite{Takahashi:2021tff} where the QCD axion with small instanton effect plays the role of a single-field inflaton. {In this paper we do not assume a setting beyond the SM, such as a small instanton effect, which would enhance the non-perturbative effect of QCD.}}

This paper is organized as follows. In the next section we discuss the basic idea of the hybrid QCD axion inflation with a generic potential of the waterfall field. In \Sec{hilltop} we {analytically} estimate the hybrid QCD axion inflation with a hilltop potential. {In \Sec{numerical}, we perform a numerical study of our inflationary model with two axion fields.} {In \Sec{searchALP}, we consider the possibility of searching {the heavier axion by accelerator experiments.}} In \Sec{reh} we study the reheating process of the two axion system. The QCD axion DM is discussed in \Sec{DM}. 
The last section is devoted to discussion and conclusions.

\section{Mixing between QCD axion and inflaton}
\label{sec:2}

\subsection{Couplings to the gluon Chern-Simons term}
Here we explain how the mixing between the QCD axion and inflaton arises.
Let us first introduce the QCD axion $a$, which dynamically solves the strong CP problem\cite{Peccei:1977hh,Peccei:1977ur,Weinberg:1977ma,Wilczek:1977pj}. 
The QCD axion has an approximate shift symmetry, called the PQ symmetry. Under the PQ symmetry transformation, the axion changes as
\beq
\laq{PQ}
a\to a+c f_a,
\eeq
where $c$ is an arbitrary real constant, and $f_a$ is the decay constant of the QCD axion. 
The QCD axion couples to gluons as
\beq
\laq{La}
{\cal L}\supset -\frac{\a_{st}}{8\pi}\frac{a}{f_a} G\tl{G},
\eeq
where $G$ and $\tilde{G}$ are the field strength of gluons and its dual, respectively, and $\alpha_{st}$ is the strong coupling constant. Thus, the PQ symmetry is anomalous and broken by QCD. One of the important parts of the PQ mechanism is the assumption that this is the only explicit breaking of the PQ symmetry. 
Through this coupling, the QCD axion acquires a temperature-dependent potential,
\beq
\laq{pot}
V_{\rm QCD}(a)=\chi(T)\(1-\cos{\(\frac{a}{f_a}\)} \),
\eeq
where $\chi{(T)}$ is the topological susceptibility. At low energy the potential form deviates from the simple cosine function, but  this approximation is sufficient for our purposes. 
The temperature dependence of $\chi(T)$ has been evaluated by several groups using the lattice QCD calculations~\cite{Berkowitz:2015aua,Bonati:2015vqz,Petreczky:2016vrs,Borsanyi:2016ksw,Frison:2016vuc,Taniguchi:2016tjc}.
According to Ref.~\cite{Borsanyi:2016ksw}, it is given by
\begin{align}
\chi(T)=\begin{cases}
 \displaystyle{\chi_0 \left(\frac{T_{\rm QCD}}{T}\right)^{2n_{\rm QCD}} } &T \gtrsim T_{\rm QCD}\vspace{3mm}\\
\displaystyle{\chi_0   }& T\lesssim  T_{\rm QCD}
 \end{cases}
\end{align}
 with $\chi_0\simeq (75.6\MEV)^4$, $T_{\rm QCD}\simeq 153\MEV$, and $n_{\rm QCD}\simeq 4.08$.
The strong CP problem is then dynamically solved {in the vacuum,} because the axion is stabilized at the potential minimum, $a=0$, where CP is conserved. 
Here we  implicitly assume that any other heavy degrees of freedom coupled to the gluon Chern-Simons term are integrated out. However, as we shall see, this is not necessarily justified in the early universe.

Let us introduce an axion-like particle (ALP) $\phi$, which is coupled to gluons in a similar way to the QCD axion,
\beq
\laq{Lag}
{\cal L}\supset \frac{\a_{st}}{8\pi}\(-n_{\rm mix} \frac{\phi}{f_\phi}+\frac{a}{f_a}\)G\tl{G},
\eeq
where $f_\phi$ is the decay constant of the ALP, and $n_{\rm mix}$ is an 
integer. We also assume that the ALP has its own potential $V_\phi(\phi)$ from other sources 
such as couplings to hidden Abelian gauge sectors. Then, the total potential is given by\footnote{
Our argument is actually more general since the potential can be rewritten in a slightly different form by field redefinition. In particular, 
one could extend the analysis to a general inflaton coupled to QCD. In this case, the QCD axion may be mixed with a certain combination of the inflaton field, but it can be analyzed in the same way.
}
\beq
V(a,\phi)=V_{\rm QCD}(a-r_{\rm mix}\phi)+V_\phi(\phi),
\eeq
where we have defined $r_{\rm mix}$ as 
\beq 
r_{\rm mix}\equiv n_{\rm mix} \frac{f_a}{f_\f}
\laq{k}.
\eeq
We can see that the QCD axion and the ALP mix through the first term at energy scales below the QCD scale. Note that the above potential does not spoil the PQ mechanism. If $\phi$ is much heavier than the QCD axion in the vacuum, we can integrate out $\phi$, and it is $a$ that plays a role of the QCD axion. As we shall see shortly, the mass eigenstates during inflation are given as a mixture of $a$ and $\phi$, but when one of them is the major component, we will loosely call it QCD axion or ALP.

Now, for this mixing to be effective during inflation, the Hubble parameter must be below the QCD scale:
\beq
\laq{cond3}
H_{\rm inf}<\Lambda_{\rm QCD},
\eeq
where $H_{\rm inf}$ is the Hubble parameter during inflation.
This is the assumption made in this paper.\footnote{The assumption can be relaxed if the QCD scale is enhanced  during the inflation in some way~\cite{Dvali:1994ms,Jeong:2013xta}.
} Furthermore, as we will see in the next subsection, we can explain the observed spectral index by considering axion hybrid inflation. This requires that $r_{\rm mix}\gg 1$, which is assumed below. This assumption also implies that the inflaton is strongly coupled to the SM sector, and the reheating will be very efficient. Such an efficient reheating is necessary when the inflation scale is low and the inflaton mass is relatively light.

\subsection{
Single-field description and inflationary path
}

We are ready to discuss the mechanism of the  hybrid inflation with the QCD axion. 
In order to have slow-roll inflation, there must be at least one flat direction in the potential $V(a, \phi)$, which satisfies the slow-roll conditions. Furthermore, for the inflationary trajectory to be well defined in the early stages of inflation, it is necessary to consider a situation where one of the mass eigenvalues at a point is lighter than $H_{\rm inf}$ and the other is heavier than $H_{\rm inf}$. In this case, the heavy direction will stabilize at its potential minimum and the inflaton will slow-roll along the light direction.

There are two main possibilities as to which of the potentials, $V_{\rm QCD}$ or $V_\phi$, contributes to the mass in the heavy direction. If $V_\phi$ is the main contributor to the heavier mass, then integrating $\phi$ during inflation would effectively make the QCD axion $a$ in the lighter direction, but the QCD axion potential \eq{pot} would not cause inflation unless $f_a$ exceeds the Planck scale. Thus, we consider here the situation where $V_{\rm QCD}$ gives the main contribution to the mass in the heavy direction.
Thus, the curvature of the potential must satisfy
\begin{align}
\laq{cond}
 |V''_\phi(\f_{0})| < H_{\rm inf}^2< (r_{\rm mix}^2+1) V''_{\rm QCD}(a_{0}-r_{\rm mix}\f_{0}), 
\end{align}
where $a_{0}$ and $\phi_{0}$ are the reference field values in early stages of inflation, well before the horizon exist of the CMB scales. 
 In this case, it is $V_\phi$ that has the necessary potential shape for inflation to occur, especially the flat part of the potential.\footnote{
The effective potential of the inflaton is {slightly different from $V_\phi$ due to the mixing effect.} See \Eq{eff}. One can see that $V_\phi$ must include a flat region for slow-roll inflation. 
 } In other words, the modulus of its first derivative, $|V'_\phi|$, must be sufficiently small at $\phi = \phi_0$. For example, in the case of the hilltop inflation, it is near the potential maximum {of $V_\phi$}, and in the case of the inflection point inflation, it is near the inflection point {of $V_\phi$}. Setting the initial condition for $a$ and $\phi$ around $a_0$ and $\phi_0$ can be thought of as fine tuning, which is a common initial value problem for low-scale inflation. This problem could be justified by a measure for eternal inflation or by some other mechanism. One example is discussed in Appendix \ref{app:2}.

Let us find the inflationary path along the light direction
by neglecting the contribution from $V''_\f$. The heavy and light eigenstates of the mass matrix at the reference point, denoted by $A_H$ and $A_L$, respectively, are given by
\begin{align}
A_H & \simeq \frac{1}{\sqrt{1+r_{\rm mix}^2}} \left(-\hat a+r_{\rm mix}\hat \f  \right) \simeq \hat \phi,
\label{AH}\\
\label{AL}
A_L & \simeq \frac{1}{\sqrt{1+r_{\rm mix}^2}} \left(r_{\rm mix}\hat a + \hat \f \right)\simeq \hat a,
\end{align}
where we have defined $\hat a \equiv a-a_{0}$ and $\hat \phi= \phi-\phi_{0}$, and
 we have used $r_{\rm mix}\gg1$ in the last equality.
The heavy $A_H$ has a positive and large mass squared from the QCD potential,  
\beq
\label{eq:massH}
M_H^2 \simeq (1+r_{\rm mix}^2) V_{\rm QCD}''.
\eeq
It is stabilized around the local minimum, determined by
\beq
\laq{cond2}
\frac{\partial V}{\partial A_H}=\frac{r_{\rm mix}}{\sqrt{1+r_{\rm mix}^2}} V'_\phi(\f_{0})-\sqrt{r_{\rm mix}^2+1}V_{\rm QCD}'(a_{0}-r_{\rm mix}\f_{0}) = 0.
\eeq
The reference field values $a_0$ and $\phi_0$ must be chosen so that this condition is met.
On the other hand, $A_L$ has a much lighter mass squared, $M^2_L=\O(V_\f''/r_{\rm mix}^2)$, and can be identified with the inflaton.

When $M_H^2 \gg H_{\rm inf}^2$, we can integrate out $A_H$
to obtain the low-energy effective potential for $A_L$,
\beq
\laq{eff}
V_{\rm eff}(A_L)\;=\; V_\phi\left( \f_{0}+\frac{1}{\sqrt{1+r_{\rm mix}^2}} A_L\right)- \left.\frac{r_{\rm mix}^2}{1+r_{\rm mix}^2}\frac{ (V'_\f)^2 }{2 M_H^2}\right|_{A_H=0}+\cdots.
\eeq
The second term comes from the mixing between $A_L$ and $A_H$ in $V_\f$, and the dots represent higher order corrections. We can neglect the second term around $A_L\approx 0$ in the equation of motion  for $A_L$, since its contribution is suppressed by a factor of  $V_\f''/r_{\rm mix}^2V_{\rm QCD}''$ compared to the first term.  
Since $A_L\approx a$ for $r_{\rm mix}\gg 1$, the inflaton is mostly the QCD axion. Note that its potential is determined by $V_\phi$, not by the non-perturbative QCD effects.   This is because $V_\phi(\phi)$ is sufficiently flat at $\phi \approx \phi_0$, and $A_L$ feels the potential from $V_\phi$ via a level crossing. It is also analogous to the alignment mechanism~\cite{Kim:2004rp}. Thus,  the inflation is driven by the QCD axion through the mixing.

Now let us assume that the QCD axion, or $A_L$,  drives inflation with the potential $V_{\rm eff}$. For this to happen,
the slow-roll conditions,  $|\eta| < 1, \varepsilon < 1$, must be satisfied, where the slow-roll parameters are defined by
\begin{align}
\varepsilon & \equiv
\frac{M_{\rm pl}^2}{2} \left(\frac{V'}{V}\right)^2, \laq{slowroll1}\\
\eta  & \equiv
M_{\rm pl}^2 \frac{V''}{V} \laq{slowroll2}
\end{align}
with the inflaton potential $V$.
Substituting \Eq{eff} into the above, one can see that the slow-roll parameters scale as
$
\varepsilon,\eta \propto ({1+r_{\rm mix}^2})^{-1/2}.
$
Thus, the suppression can be significant for large $r_{\rm mix}$.
If the $\varepsilon$ parameter is sufficiently small, eternal inflation could take place; the condition for the eternal inflation is given by $H_{\rm inf}^3\gtrsim 2\pi V_{\rm eff}'=\O(V_\f'/r_{\rm mix})$. 

So far we have discussed a setup in which the QCD axion  drives the inflation at early stages before the horizon exit of the CMB scales. There are two main possibilities as to where inflation ends:
\begin{itemize}
\item The inflation path follows the direction of the QCD axion all the way to the end of inflation. At some point the ALP becomes a waterfall field and inflation ends.
\item At some point, the inflation path changes from the direction of the QCD axion to the direction of the ALP, and inflation continues for a while.
\end{itemize}
The first case can be further divided into two ways of ending inflation: when inflation ends before the waterfall by the ALP field, or when inflation ends as the ALP field waterfalls.  We call the former {\it a single-field regime} and the latter {\it an instantaneous waterfall regime}. In both cases, the dynamics during inflation can be described using the single-field approximation. {The single-field regime} was studied in Ref.~\cite{Takahashi:2019pqf} and is reviewed in the next section. The parameter region derived from this regime is in tension with the stellar cooling limit of the QCD axion.

In the second case, we must carefully consider the mixing of the two fields to describe inflationary dynamics. We call this case {\it a prolonged waterfall regime}. This, as well as the instantaneous waterfall regime, has not been studied before and is the focus of this paper. We will show that taking the two-field dynamics into account allows for a larger decay constant of the QCD axion consistent with the stellar cooling limits and gives a new production mechanism of the QCD axion DM with $f_a\sim 10^{9-10}\,\GEV$.

\section{QCD axion hybrid inflation}
\lac{hilltop}
In this section, we first review the single-field regime of QCD axion hybrid inflation based on Ref.~\cite{Takahashi:2019pqf}, and derive the limit on $f_a$.  Then we discuss the case of the waterfall regimes. A detailed analysis of the viable parameter region requires numerical calculations, which will be discussed in the next section.

\subsection{Hilltop potential for ALP}
\lac{rev}
Let us consider a concrete potential of $\f$ with a periodicity of $2 \pi f_\phi$\footnote{Our mechanism works even if $\f$ is not the axion-like particle, as long as it couples to the Chern-Simons term.  Since we use the quartic potential \eq{app} to describe the inflation, any UV model with the quartic hilltop potential will also work. }.
If it is dominated by a single cosine term such as $V_\f =\L^4(1-\cos{(\f/f_\f)}),$ the potential height $\sim \L^4$ must be less than $\chi_0$ to satisfy both \Eqs{cond} and \eq{cond2} for $n_{\rm mix}=\O(1)$. Such a low-scale inflation is problematic at least for reheating.    
In addition, a super-Planckian $f_a$ is needed to drive the inflation.

Instead, we focus on a potential dominated by multiple cosine terms. 
Such periodic potentials have been considered in the single-field inflation model under the name of multi-natural inflation~\cite{Czerny:2014wza, Czerny:2014xja,Czerny:2014qqa,Higaki:2014sja}, 
where multiple cosine terms conspire to realize a sufficiently flat potential. 
For simplicity, let us use two cosine terms to express the potential, 
 \begin{align}
\label{eq:DIV} 
V_{\f}(\phi) = \Lambda^4\(\cos\(\frac{\phi}{f_\f} + \Theta \)- \frac{\kappa }{n^2}\cos\(n\frac{\f}{f_\f }\)\)+{\rm 
const.},
\end{align}
where $n$ ($>1$) is an even integer, $(\kappa-1)$ and $\Theta$ parameterize the relative magnitude and phase of the two terms, respectively,  and the last constant term is introduced to make the cosmological constant vanishingly small in the present vacuum.\footnote{\label{f1}One can consider a case where the first cosine term in \Eq{DIV} contains another positive integer $n'_{\rm inf}${, satisfying $n'_{\rm inf} < n$}. It is easy to extend our analysis to this case by redefining the decay constant and allowing $n$ and $n_{\rm mix}$ to be fractional numbers. }   $\k-1$ and $\Theta$ can be non-zero if the two terms come from different sources, and their typical value depends on the UV completion. 

The flat-top potential in the axion inflation has several possible UV origins, e.g. in supergravity\cite{Czerny:2014xja,Czerny:2014qqa,Higaki:2014sja}, a  component of the Higgs fields in the D-flat direction of the MSSM~\cite{Murai:2023gkv}, and extra dimensions~\cite{Croon:2014dma}. A similar potential with an elliptic function is also obtained in the low-energy limit of string-inspired setups~\cite{Higaki:2015kta, Higaki:2016ydn}. In this paper, unless otherwise stated, we take \beq\laq{hilltopcond} \Theta=0, ~~\k=1.\eeq   This particular setup without the CP phase can be realized in Ref.~\cite{Croon:2014dma}. Similar potential can also be found in Refs.~\cite{Higaki:2015kta, Higaki:2016ydn}.

The inflation happens at around $a_{0}= 0 \AND \f_{0}= 0,$ where \Eqs{cond} and \eq{cond2} are satisfied. 
We consider the field space of $\f>0$ and $a>0$ without loss of generality. 
Let us expand the potential around $\f=0$, 
\beq
\label{eq:app}
V_{ \f}(\phi) \simeq V_0   - \lambda \f^4+\cdots,    
\eeq
where the dots represent terms with negligible effects on the inflaton dynamics during  inflation, and $V_0$ and $\lambda$ are given by
 \begin{align}
 \label{eq:V0}
 V_0 & \equiv \beta_n \Lambda^4,\\
  \lambda &\equiv \frac{n^2-1 }{4!}\(\frac{\Lambda}{f_\f}\)^4
 \label{eq:lambda}
\end{align}
with
$$\beta_n=2 - \frac{2}{n^2} \sin^2{\frac{n \pi}{2}}.$$
Here we have chosen $V_0$ so that the potential vanishes at the minimum.

After inflation,  $\phi$ is stabilized at one of the potential minima of $V_\phi$, 
\beq
\label{phimin}
\phi_{\rm min} = \pi f_\phi.
\eeq
As we shall see below, the dynamics of the QCD axion $a$ is limited around the origin during and after inflation. As $\phi$ moves toward $\phi_{\rm min}$ after the end of inflation, the potential for $a$ shifts by an amount of $n_{\rm mix} \pi$~\cite{Takahashi:2019pqf}. 
When the cosmic temperature becomes comparable to the
QCD scale, $a$ starts to oscillate about the nearest minimum,
\beq
\label{amin}
a_{\rm min} = \left\{
\begin{array}{cc}
  \pi f_a   &  ~~{\rm for~~}n_{\rm mix}~ = {\rm ~ odd} \\
   0  &  ~~{\rm for~~}n_{\rm mix}~ = {\rm ~ even} 
\end{array}
\right.
\eeq
with an initial amplitude determined by the inflaton dynamics.
Later we will estimate the initial oscillation amplitude, which is important to determine the QCD axion abundance.

\subsection{Single-field regime in a corner}
Following Ref.~\cite{Takahashi:2019pqf}, we derive an upper bound on $f_a$ in the single-field regime of the QCD axion hybrid inflation. This upper bound is outside the so-called QCD axion window, and thus the single-field regime is in tension with the stellar cooling bounds.

At $a\approx 0$ and $\f\approx0$, we can integrate out $A_H$ to obtain the effective potential \eq{eff} for $A_L$, as discussed in the previous section. The effective potential is given by
\beq
\laq{Veff}
V_{\rm eff} = V_0-\lambda_{\rm eff} A_L^4-  \frac{8r_{\rm mix}^2\l_{\rm eff}^2 f_a^2 }{(1+r_{\rm mix}^2)\chi_0}A_L^6+\cdots, 
\eeq
where we have used $M_H^2\simeq (1+r_{\rm mix}^2)\chi_0/f_a^2,$ and defined
\beq
\lambda_{\rm eff}\equiv \frac{\l}{(1+r_{\rm mix}^2)^2}\approx \frac{n^2-1 }{4!}\(\frac{\Lambda}{n_{\rm mix} f_a}\)^4.
\eeq 
Notice that $V_{\rm eff}$ is not sensitive to $f_\f$ when $r_{\rm mix}\gg 1$.
The field range in which this effective potential is valid is given
by
\beq
\laq{quartic}
A_L\lesssim A_L^{\rm cutoff}\equiv\frac{1}{ \sqrt{12 \lambda_{\rm eff}}}\frac{\sqrt{\chi_0}}{f_a}, 
\eeq
where we have used $r_{\rm mix}\gg1.$ Within this range, the sixth-order term in the effective potential is negligible compared to the quartic term in the equation of motion for $A_L$. On the other hand, when $A_L \sim A_L^{\rm cutoff}$,   the sixth-order term and the higher-order terms become as large as the quartic terms. At this point, effective theory breaks down, and we need to consider the two-field dynamics.

If the universe is dominated by $V_0$, a quartic hilltop inflation could take place around $A_L \simeq 0$. For this to happen, the slow-roll parameters must be less than unity, $|\eta| \lesssim 1$ and $\varepsilon\lesssim 1$.  The inflation ends when one of these parameters equals unity. In the low-scale inflation, $\varepsilon$ is much smaller than {$|\eta|$} during inflation, and it is {$|\eta|$} that first becomes equal to unity. 
 Thus, the field value of $A_L$ at the end of inflation is given by
\beq
\laq{infend}
A_{L}^{\rm end}\simeq \sqrt{\frac{2\b_n}{n^2-1}}\frac{n_{\rm mix}^2f_a^2}{M_{\rm pl}}.
\eeq
In the single-field regime, inflation ends within the validity of the effective potential, i.e. $A_L$ becomes equal to $A_{L}^{\rm end}$ before it reaches $A_L^{\rm cutoff}$, 
\beq
\laq{QCDinflation}
A_{L}^{\rm end} \lesssim A_{L}^{\rm cutoff}.
\eeq
From this we obtain an upper bound on $f_a$, 
\beq
f_a\lesssim  2\times 10^7\GEV \cdot \(n^2-1\)^{1/6}\b_n^{-1/6}n_{\rm mix}^{-2/3} \(\frac{10^{-13}}{\lambda_{\rm eff}}\)^{1/6},
\eeq
where we adopt the typical value of the quartic coupling to explain the CMB normalization.
However, due to the neutrino burst duration of SN1987A~\cite{Mayle:1987as,Raffelt:1987yt,Turner:1987by,Chang:2018rso} and
the cooling of neutron stars~\cite{Hamaguchi:2018oqw}, $f_a\gtrsim 10^8\GEV$ is required.\footnote{
This astrophysical bound can be relaxed by considering a hadrophobic/astrophobic QCD axion~\cite{DiLuzio:2017ogq, Bjorkeroth:2019jtx, DiLuzio:2022tyc}, which can  be UV completed in grand unified theory~\cite{Takahashi:2023vhv}.} 
To satisfy the astrophysical bounds, the system must leave the validity of the effective potential before the end of the inflation. This is the case when the higher dimensional term in \Eq{Veff} becomes relevant towards the end of the inflation.

For now, let us ignore the astrophysical limit for $f_a$ and calculate the predictions of this model for curvature perturbations. 
The  power spectrum of the curvature perturbation is given by
\beq
\label{eq:pln}
\D^2_{\mathcal{R}}(k) 
=\frac{H_{\rm inf}^2}{8\pi^2  \varepsilon M_{\rm pl}^2},
\eeq 
where the right hand side is evaluated at the horizon exit of the comoving mode $k$. The observed CMB normalization is 
\beq
\label{pn}
\D^{2}_{\mathcal{R}}(k_*) \simeq 2.1 \times 10^{-9}
\eeq
at the pivot scale $k_* = 0.05\, {\rm Mpc}^{-1}$\cite{Akrami:2018odb}.
Here the scale factor is set to be unity at present, and
we use ``$*$" to denote the quantity at the horizon exit of the CMB scale. This formula (\ref{eq:pln}) is applicable when the inflaton dynamics are well described by a single-field slow-roll inflation with a canonical kinetic term. 
In other words, any heavy degrees of freedom can be integrated out, and they are not excited during inflation.

The formula for the scalar spectral index $n_s$ is given 
by
\beq
\label{eq:nsform}
n_s(k) \equiv 1+ \frac{d \ln \D^2_{\mathcal{R}}(k)}{d \ln k} \;\simeq\; 1- 6 \varepsilon + 2\eta.
\eeq
In low-scale inflation, $\varepsilon$ is much smaller than $|\eta|$, and it is further simplied to $n_s \approx 1+2\eta$. 
The observed value of $n_s$ from CMB observations is~\cite{Akrami:2018odb}
\beq
\label{nsCMB}
n_s(k_*)= 0.965 \pm 0.004,
\eeq
where we have adopted the result of the $\it Planck$ TT, TE, EE + lowE.

The number of e-folds, $N_*$, corresponding to the time between  the horizon exit of CMB scales, $t=t_*$, and the end of inflation, $t=t_{\rm end}$, can be calculated as $N_*= \int_{t_*}^{t_{\rm end}}{H_{\rm inf} \,dt}$. 
This is also related to the thermal history after inflation. If the reheating of the universe is completed soon after the inflation, one obtains
\beq
\label{eq:efold}
N_*\simeq 28 -\log{\(\frac{k_* }{0.05\, {\rm Mpc^{-1}}}\)}+\log{\(\frac{V_0^{1/4} }{10\,\TEV}\)}.
\eeq
Since we are considering low-scale inflation with $H_{\rm inf} < \Lambda_{\rm QCD}$, the number of e-folds is much smaller than the typical value for high-scale inflation. Specifically, we have $N_*\lesssim 40$ for $H_{\rm inf}\lesssim 1\GEV$.

In the single-field regime, the inflaton dynamics is the same as quartic hilltop inflation.  For the potential given by the first two terms of \eq{Veff}, we obtain (see e.g. Refs.~\cite{Daido:2017wwb,Daido:2017tbr,Takahashi:2019qmh})
\begin{align}
\frac{\L}{n_{\rm mix} f_a}&\;\simeq\;   1.2\times 10^{-3 }\(\frac{3}{n^2-1}\)^{\frac{1}{4}}\(\frac{30}{ N_*}\)^{\frac{3}{4}},\\
n_s&\;\simeq\; 1-\frac{3}{N_*}.
\end{align}
Here, the CMB normalization (\ref{pn}) fixes the quartic coupling $\lambda_{\rm eff}$ as a function of the number of e-folds. Since the number of e-folds is relatively small, the predicted spectral index is bounded above as $n_s \lesssim 0.93$, which is too small to explain the observed value (\ref{nsCMB}).\footnote{The predicted spectral index can be increased and the tension can be relaxed by introducing a non-vanishing $\Theta$ and $\kappa-1$~\cite{Takahashi:2013cxa,Takahashi:2019pqf}. We do not pursue this possibility in this paper.}

Therefore, the single-field regime of QCD axion hybrid inflation 
is disfavored from the point of view of both stellar cooling limits and CMB observations. This leads us to consider the other two waterfall regimes.

\subsection{Waterfall regimes of axion hybrid inflation}

Here we relax the condition~\eq{QCDinflation}, and consider the opposite case,
\beq
\laq{hybQCDinf}
A_L^{\rm end} \gtrsim A_{L}^{\rm cutoff},
\eeq 
which implies $f_a\gtrsim 10^{8}\GEV$.
Interestingly, as we will see in the rest of the paper, in this case the axion decay constant is significantly increased, and we also have a better fit to the CMB data. In addition the successful production of QCD axion DM is possible with $f_a\simeq 10^{9}$\,GeV.

In the case of \eq{hybQCDinf}, the light mass eigenstate $A_L$ will exceed $A_L^{\rm cutoff}$ at a certain point, and then,  it becomes necessary to consider the dynamics of the two fields. 
Two main cases are possible. Initially, the inflation is primarily driven by  the inflaton in the direction of the QCD axion, but as the slope of $V_{\phi}$ gradually increases, the dynamics in the ALP direction become more significant. If $f_{\phi}$ is somewhat small, the waterfall will occur immediately after the inflaton starts to move in the ALP direction, ending the inflation. This is the immediate waterfall regime. On the other hand, if $f_{\phi}$ is large enough, the inflation will continue in the ALP direction for some time. This is the prolonged waterfall regime. Each of these regimes is discussed below.

In the instantaneous waterfall regime with a relatively small $f_\phi$, the spectral index increases compared to the  quartic hilltop inflation in the single-field regime.
This can be understood as follows. When $A_L$ reaches $A_L^{\rm cutoff}$, one can no longer integrate out the $A_H$, and $A_H$ starts to move rapidly toward its potential minimum, ending the inflation. 
Thus the end of inflation is earlier
than the case of the quartic hilltop inflation. In other words, the field value of $A_L$ at the horizon exit of the CMB scales becomes smaller. Since $|\eta|\propto \lambda_{\rm eff} A_L^2$ becomes smaller, $n_s \simeq  1-2|\eta|$ becomes larger. Also, since the first derivative of the inflaton potential is small throughout the inflation, the value of the inflaton field does not change significantly compared to the single-field regime.

The power spectrum of the curvature perturbation scales as $P_{\cal R}\propto \L^4 f_a^8/A^6_L$, and so it results in a smaller $\L$. Assuming that the inflation ends instantaneously when $A_L\sim A_L^{\rm cutoff}$ and the field value at the horizon exit of the CMB scales is not very different from  $A_L^{\rm cutoff}$, the CMB normalization gives
\beq
\L\sim 30\TEV \(\frac{10^9\GEV}{ f_a}\)^{1/8}\(\frac{3}{n^2-1}\)^{1/16}.
\eeq
This is consistent with the numerical result to be discussed later.
Although we need a numerical study of the two-field dynamics to estimate the precise inflation scale, we can still analytically estimate the curvature perturbation as in the single-field regime.

For larger $f_\phi$, the ALP can drive inflation after $A_L$ reaches $A_L^{\rm cutoff}$. In this prolonged waterfall regime, it is more subtle to explain the observed spectral index. In fact, both $a$ and $\phi$ could make comparable contributions to the curvature perturbations, and thus to the spectral index. Then, the predicted spectral index can be enhanced to match the observations. For this to happen, the heavy eigenstate $A_H$ should have a mass comparable to {or slightly smaller than} the Hubble parameter,
\beq
\laq{twofield}
M_H\lesssim H_{\rm inf},
\eeq
around the horizon exit of the CMB scales, so that it acquires unsuppressed fluctuations at superhorizon scales. This requires $f_a \gtrsim 10^8\GEV$. To estimate the curvature perturbation as well as the spectral index in this regime, we must rely on the numerical calculations. This is the topic of the next section.

\section{Numerical analysis } 
\lac{numerical}

In the previous section, it was found necessary to consider the waterfall regime of the QCD axion hybrid inflation in order to satisfy the stellar cooling limits of the QCD axion. In this case, the mixing of the two fields evolves non-trivially in time during the inflation. Therefore, a detailed numerical analysis is required. In Appendix \ref{app:nc} we summarize the general formulation for multiple light scalar fields necessary for numerical calculations of the curvature perturbations, etc.

The numerical results presented in this section are obtained from the following numerical calculations. 
We first solve \Eq{backgroundfield} with an initial condition of $\bar{\f}_i\approx 0$ and $\bar \f_i' = - M_{\rm pl}^2 \partial_{\bar \f_i}\pqty{\log{V}}$ for the homogeneous background component, $\bar{\phi}_i$. Here the bar represents the homogeneous component, and the prime denotes the derivative with respect to the e-folds, {and $i = 1, 2$ denotes the QCD axion and ALP.}
Throughout this numerical calculations, we set $n=2$ and $n_{\rm mix}=1$.
Then we solve {the equations of motion for the fluctuations,} \Eqs{pEoM} and \eq{Beq1},  with the initial condition of the Bunch-Davies vacuum.
For each mode, the initial time for the numerical calculations is chosen so that the mode is initially well within the Hubble horizon.

There are two cases in the waterfall regime. One is the instantaneous waterfall regime and the other is the prolonged waterfall regime.  They roughly correspond to $f_a\simeq 10^8\,\GEV$ and $f_a\gg 10^8\GEV$, respectively. In \Sec{single} and \Sec{multi} we first study the inflaton dynamics and the curvature perturbations for the respective sample points. We then discuss the viable parameter region by comparing it to the CMB data in \Sec{parameterregion}.

\begin{figure}[!t]
\begin{center}  
          \includegraphics[width=0.965\textwidth]{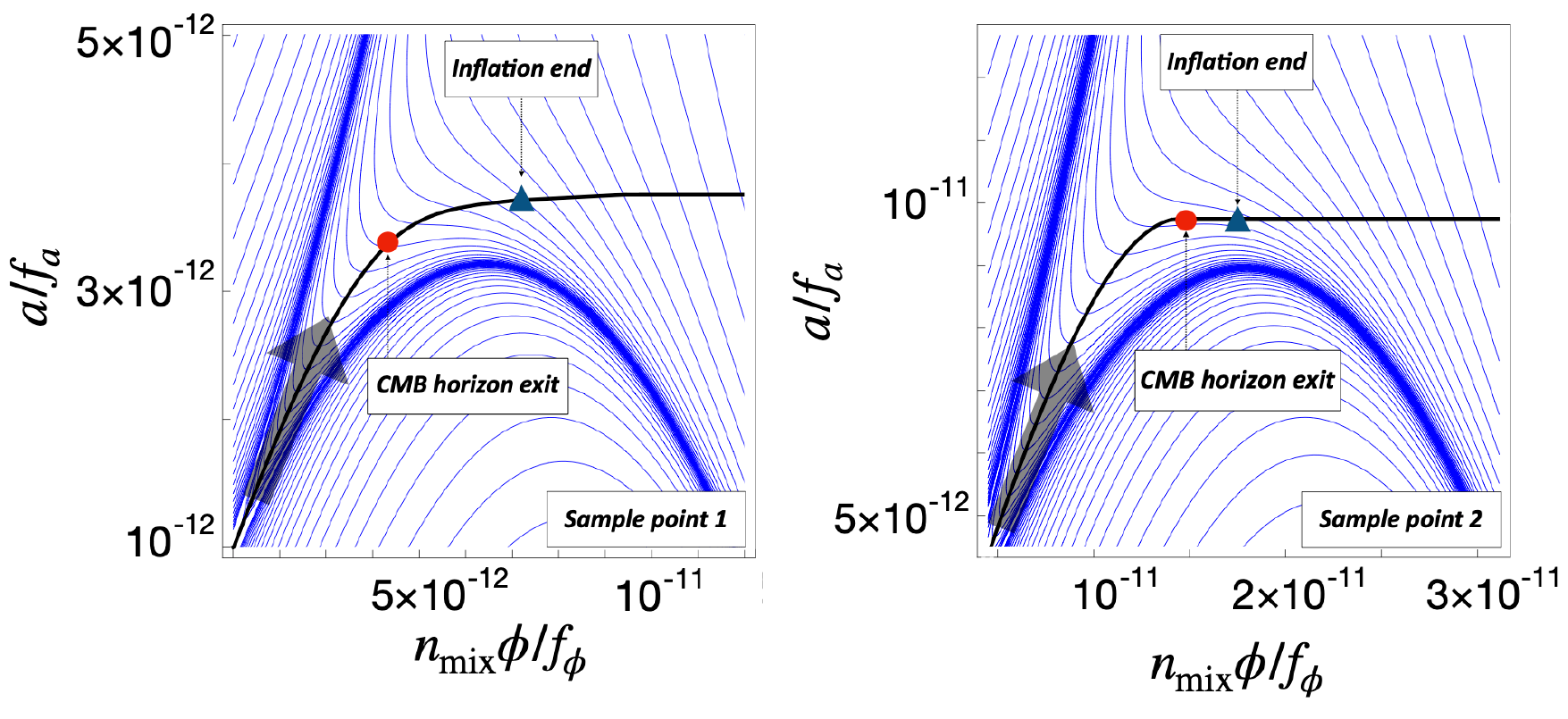}
      \end{center}
\caption{The inflaton trajectories  during the QCD axion hybrid inflation corresponding to the sample points 1 (left panel) and 2 (right panel)  in the instantaneous and prolonged waterfall regimes, respectively.  In both cases, the inflaton trajectory is well defined
 because it is effectively single-field inflation in the early stages.
 Also plotted in blue lines are the contours of $\log|V-V_0|$. The red point and blue triangle correspond to the horizon exit of the CMB scales and the end of inflation, respectively.
}\label{fig:2}  
\end{figure}

\subsection{Instantaneous waterfall regime
}
\lac{single}
As a sample point in the instantaneous waterfall regime, we adopt the following parameters:
\beq
[{\rm Sample~point~1}]~~~f_a \approx 1.2 \times 10^8 \GEV,\, f_\phi = 10^7 \GEV,\, 
\Lambda \approx 30 \TEV,
\eeq
where $f_a$ and $\L$ are numerically determined to be consistent with the CMB observations, and so we use $\approx$ in the above.  This is shown as sample point 1 in the left panel of Fig.~\ref{fig:nscontour}.

We show the inflaton trajectory of the background fields in the left panel of Fig.\ref{fig:2},
where the red point and blue triangle represent where the CMB scales exited the Hubble horizon and the inflation ends, respectively.\footnote{Here the end of inflation is determined by by $|\eta|=|\frac{1}{2}\frac{\dot{\varepsilon}}{H \varepsilon }|=1$. On the other hand, in the numerical calculations, we adopt $\epsilon =1$, which is a more precise definition of the end of inflation. The difference in the number of e-folds is typically $N_{\e=1} - N_{|\eta|=1} = 1-2$. 
 }
It can be seen that the horizon exit of the CMB scales is just before the trajectory begins to bend along the ALP direction and that the contour interval narrows around the blue triangle and the slope of the potential becomes steeper.
Note that since $f_a \sim 10 f_\phi$, the actual inflaton trajectory is almost along the QCD axion before the CMB horizon exit. The inflation ends soon after the trajectory changes to the ALP direction, which increases the spectral index as discussed earlier.

\begin{figure}[!t]
\begin{center}  
     \includegraphics[width=\textwidth]{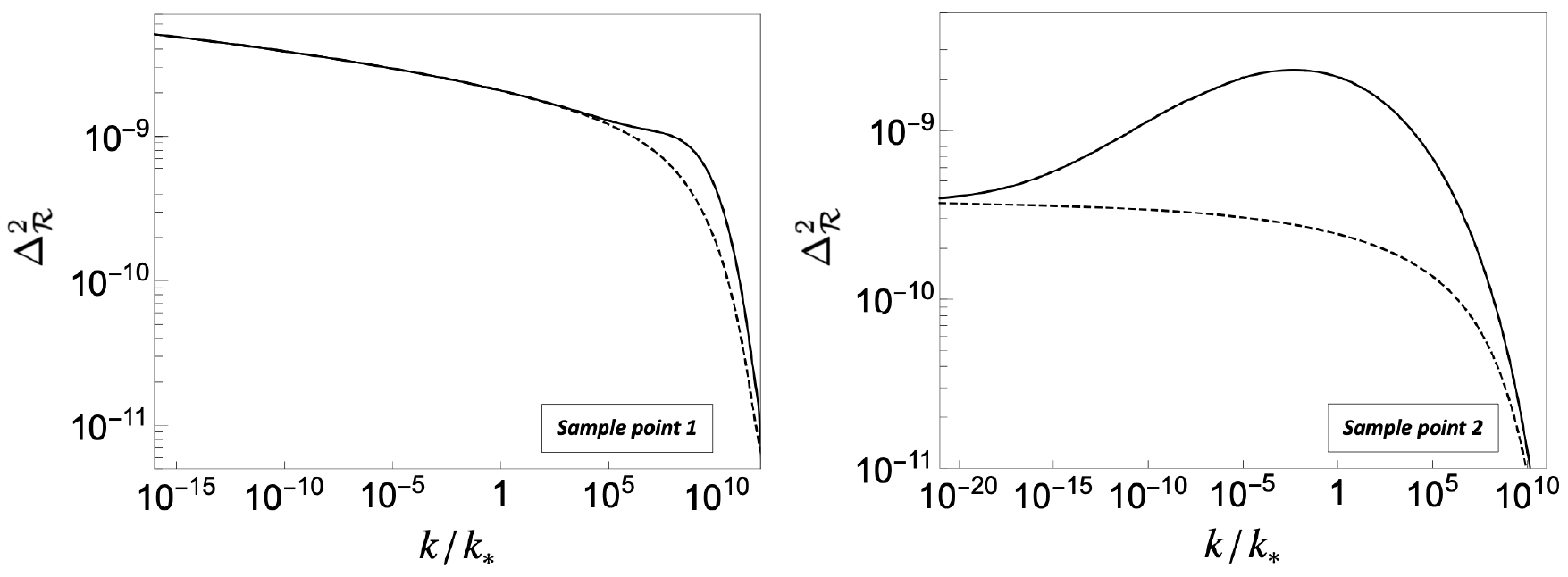}
\end{center}
\caption{The reduced power spectra of curvature perturbation $\Delta^2_{\mathcal{R}}(k)$ as a function of $k/k_*$ for the sample points. The dashed line represents the single-field approximation, while the solid line is based on the two-field analysis.
}
\label{fig:ps} 
\end{figure}

We have also calculated the reduced power spectrum of the curvature perturbation (see   \Eq{reducedpower}), which is shown as a solid line in the left panel of Fig.~\ref{fig:ps}. 
For comparison, we also show the result based on a single-field approximation (dashed line), where we only introduced the fluctuation along the background inflaton trajectory, \fpa, as the initial condition, while the perpendicular one, $\d\f_\perp$, is set to zero. 
 The mode corresponding to the CMB scale is at $k/k_*=1$. 
 As expected, we see that the curvature perturbation around the CMB scales is well described by the single-field approximation. On the other hand, at small scales we have a slight enhancement of the power spectrum compared to the single-field approximation. This is because the mass along a direction perpendicular to the inflation trajectory becomes lighter than the Hubble parameter towards the end of the inflation, and contributes to the curvature perturbation by modulating the inflation trajectory. This is a small effect in this case, but it becomes prominent in the regime we consider next.

\subsection{
Prolonged waterfall regime
}
\lac{multi}
As a sample point in the prolonged waterfall regime, we adopt the following parameters:
\beq
[{\rm Sample~point~2}]~~~f_a = 5 \times 10^9 \GEV,\, f_\phi \approx 2 \times 10^7 \GEV,\, 
\Lambda \approx 18 \TEV,
\eeq
where $f_a$ is much larger than the previous case. Here $f_\phi$ and $\L$ are numerically determined to be consistent {with} the CMB observations.
This is shown as the sample point 2 in the right panel of Fig.~\ref{fig:nscontour}.
This parameter set will also explain the QCD axion DM as we will see in Sec.\ref{chap:DM}. 

The corresponding inflaton trajectory is shown in the right panel of Fig.~\ref{fig:2}. The horizon exit of the CMB scale is around the inflaton trajectory bends to the ALP direction, where both mass eigenvalues are lighter than the Hubble parameter. One can also see that the inflationary direction is very light because the trajectory  is almost parallel to the contour lines, and thus inflation will continue. In this case, we cannot use the standard analysis for the single-field inflation to estimate the curvature perturbation, and we need  numerical calculations taking account of the fluctuations of both scalars.

We show the reduced power spectrum of the curvature perturbation (solid line) in the right panel of Fig.~\ref{fig:ps}. We can see that there is a significant deviation from the result based on the single-field approximation (dashed line), especially at the CMB scales $k/k_* \sim 1$. This implies that the effect of multiple light scalars is important to explain the CMB observations, especially the spectral index. 
On the other hand, both lines converge at both large and small wave numbers.  We can understand this behavior as follows.
For the fluctuations with small wave numbers, they exit the horizon before the trajectory bends and the contribution of \fpe is effectively suppressed because $M_H$ is  larger than $H_{\rm inf}$. As a result, only \fpa remains prominent. 
However, as the wave number gets larger, the horizon exit is  delayed and changes in the trajectory becomes important. When the trajectory bends, \fpe becomes lighter than the Hubble parameter,
and its fluctuations start to affect the curvature perturbation.
{When the fluctuations with even larger wave numbers exit the horizon,  the trajectory becomes nearly straight in the direction of the ALP. Then, while \fpe is sizable and does not decay at superhorizon scales,  it does not alter the number of e-folds $N$.
Hence, \fpe no longer contributes to the curvature perturbation.} This is similar to the usual situation where the light QCD axion acquires quantum fluctuations, but it does not contribute to the curvature perturbation. Thus, the 
power spectrum gradually approaches the single-field result. 
We will study its contribution to the isocurvature perturbation in \Sec{DM}.

In summary, our analysis revealed that the power spectrum is significantly influenced by the multi-field effect in the prolonged waterfall regime.

\subsection{Viable parameter region}
\lac{parameterregion}

We have performed the numerical calculations by varying the model parameters to delineate a viable parameter region that is consistent with the CMB data. From the resulting curvature perturbation, we can more precisely study the dependence of the spectral index $n_s$ on the decay constants $f_\f$ and $f_a$. We show the contours of the spectral index in Fig.~\ref{fig:nscontour}, where one can see the constraints imposed by the CMB data on each decay constant.  The horizontal band in the left panel corresponds to the instantaneous waterfall regime, while the vertical bands in the both panels correspond to the prolonged waterfall regime.  The vertical band in the left panel is connected to the band in the right panel. Specifically, we obtain the following bounds from the figure: 
%
\begin{figure}[!t]
    \begin{center}
        \includegraphics[width=\textwidth]{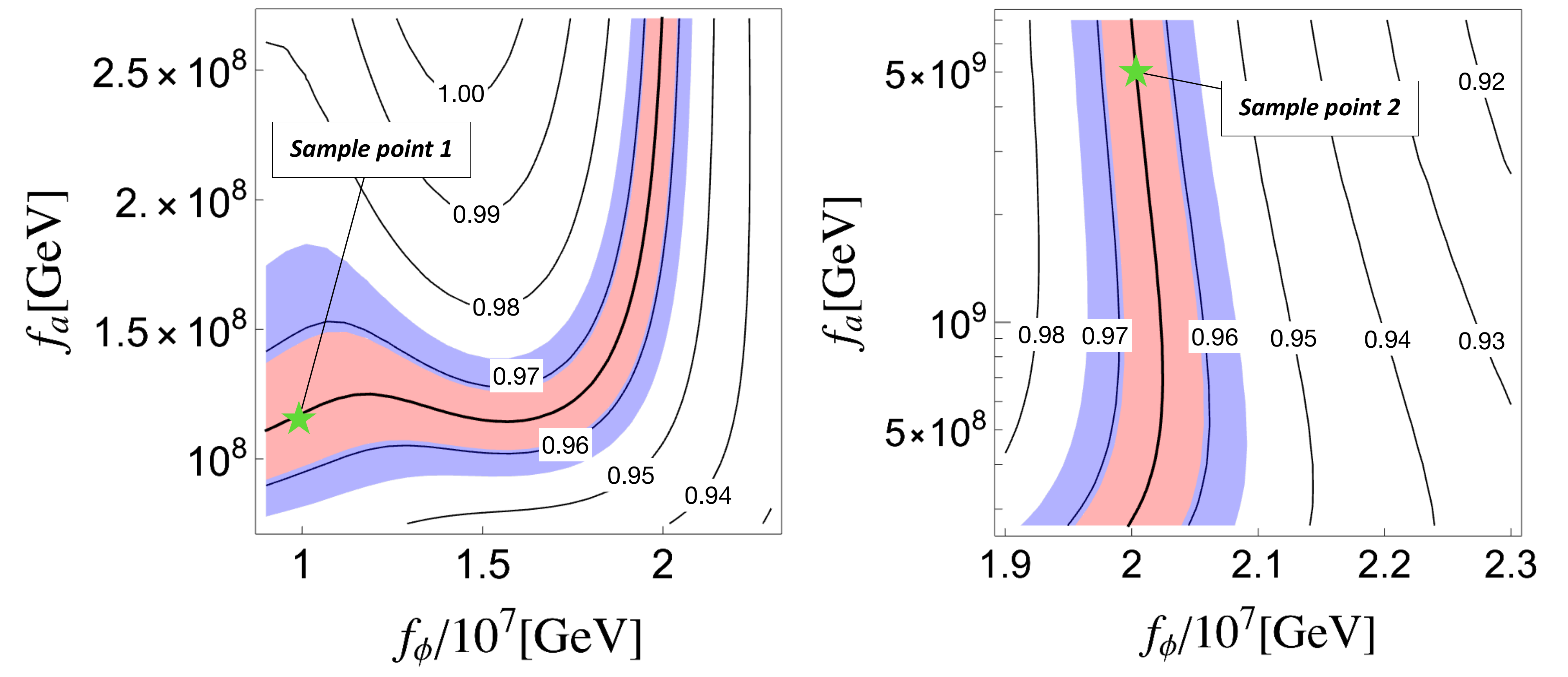}
    \end{center}
\caption{The contour plots of $n_s$ with respect to $f_\phi$ and $f_a$ in different ranges of the decay constants. The range of the spectral index $n_s$ allowed by the CMB data (\ref{nsCMB}) is shown as red ($1\s$) and blue ($2\s$) bands.
The horizontal band of the left panel corresponds to the instantaneous waterfall regime, while the vertical bands in both panels correspond to the prolonged waterfall regime.
}
\label{fig:nscontour}
\end{figure}
\begin{align}
    \label{phiup}
    f_\f &\lesssim 2.1 \times 10^7 \GEV \\
    \label{alow}
    f_a &\gtrsim 7.5 \times 10^7 \GEV.
\end{align}
Intuitively, we can understand these constraints in terms of the following two limits. The first limit is the single-field regime with $f_a \ll 10^7\GEV$. The other limit is $f_\f \gg 10^8\GEV$ and $ \L\sim 10^{-3}f_\f$, where the ALP drives a quartic hilltop inflation with the light QCD axion as a spectator. In both limits, the mixing between $a$ and $\phi$ is irrelevant during inflation. It is well known that quartic hilltop inflation with a small inflation scale (see \Eq{cond3}) predicts too low a spectral index. Therefore, both limits are not consistent with the CMB data. This argument leads us to consider a case in which the masses of the two scalars are both not too different from the Hubble parameter during the inflation in order to increase the spectral index. In fact,  $n_s$ can be greater than $1$ in the prolonged waterfall regime. This is due to the amplification of \fpe and the resulting enhancement of the curvature power spectrum for $f_a\gtrsim 3\times 10^8\GEV$ and $f_\f\simeq 2 \times 10^7\GEV$, as shown in the {right} panel of Fig.~\ref{fig:ps}.

\begin{figure}[!t]
    \begin{center}
        \includegraphics[width=\textwidth]{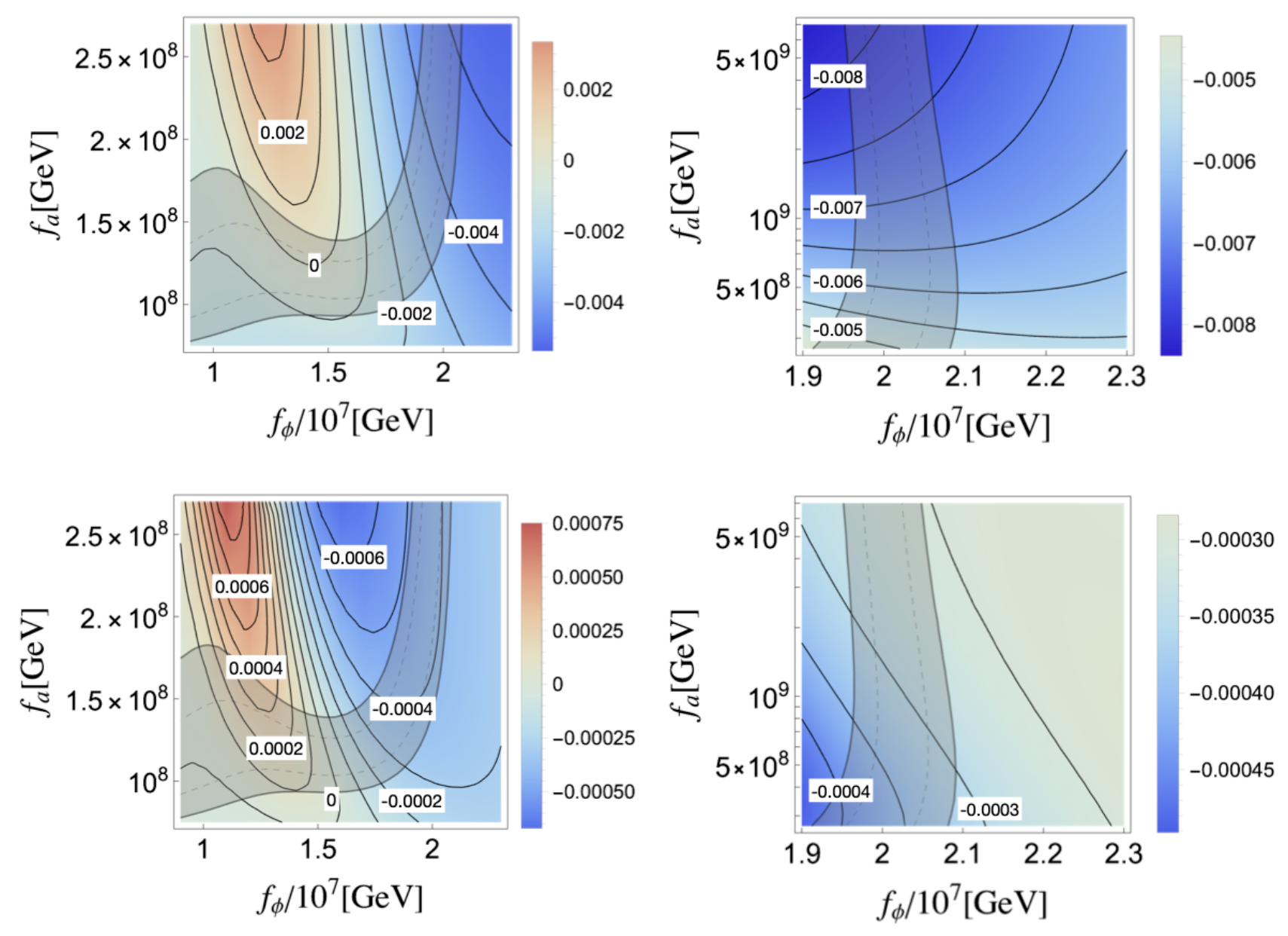}
    \end{center}
\caption{The upper panels are the contour plots of $\a_s$, while the lower panels are the contour plots of $\b_s$. We show the allowed range of $n_s$ in Fig.~\ref{fig:nscontour} by light gray bands.
}
\label{fig:runcontour} 
\end{figure}

In addition to investigating the spectral index $n_s$, we now discuss the running of the spectral index, $\a_s$, and the running of running of the spectral index $\b_s$. It is important to note that in the case of low-scale inflation, characterized by a smaller number of e-folds $N_*$ compared to high-scale inflation scenarios, the magnitudes of $\a_s$ and $\b_s$ tend to be larger \cite{Garcia-Bellido:2014gna}. Also, there is a correlation between the multi-field effect and $\a_s$, $\b_s$. As shown in the right  panel of Fig.~\ref{fig:ps}, the amplification resulting from the multi-field effect alters the shape of the curvature power spectrum, which in turn {could affect} high-order terms of $\log{\(k/k_*\)}$ in the fitting function of Eq.~\eq{PR}. We show in Fig.~\ref{fig:runcontour} the numerical results of $\a_s$ and $\b_s$. The CMB constraints on these parameters are~\cite{Akrami:2018odb}
\begin{align}
    \label{as}
    \a_s(k_*) &=-0.0045 \pm 0.0067\\
    \label{bs}
    \b_s(k_*) &=0.022 \pm 0.012,
\end{align}
where we have taken the result of Planck TT,TE,EE+lowE. One can see from  Fig.~\ref{fig:runcontour} that the calculated $\alpha_s$ and $\beta_s$ in our model are consistent with the CMB constraints,
at least where the spectral index is consistent with the CMB observation shown in the light gray regions.
A future measurement of the curvature power spectrum could test our scenario. We also note that since this is a low scale inflation, it is consistent with the non-observation of the tensor mode. As a result, our scenario can successfully explain the CMB data,
and in particular it provides a correlation between the parameters $f_a\AND f_\phi$ in the viable parameter region.

\section{Possible experimental searches for ALP}
\lac{searchALP}

\begin{figure}[!t]
\begin{center}  
     \includegraphics[width=\textwidth]{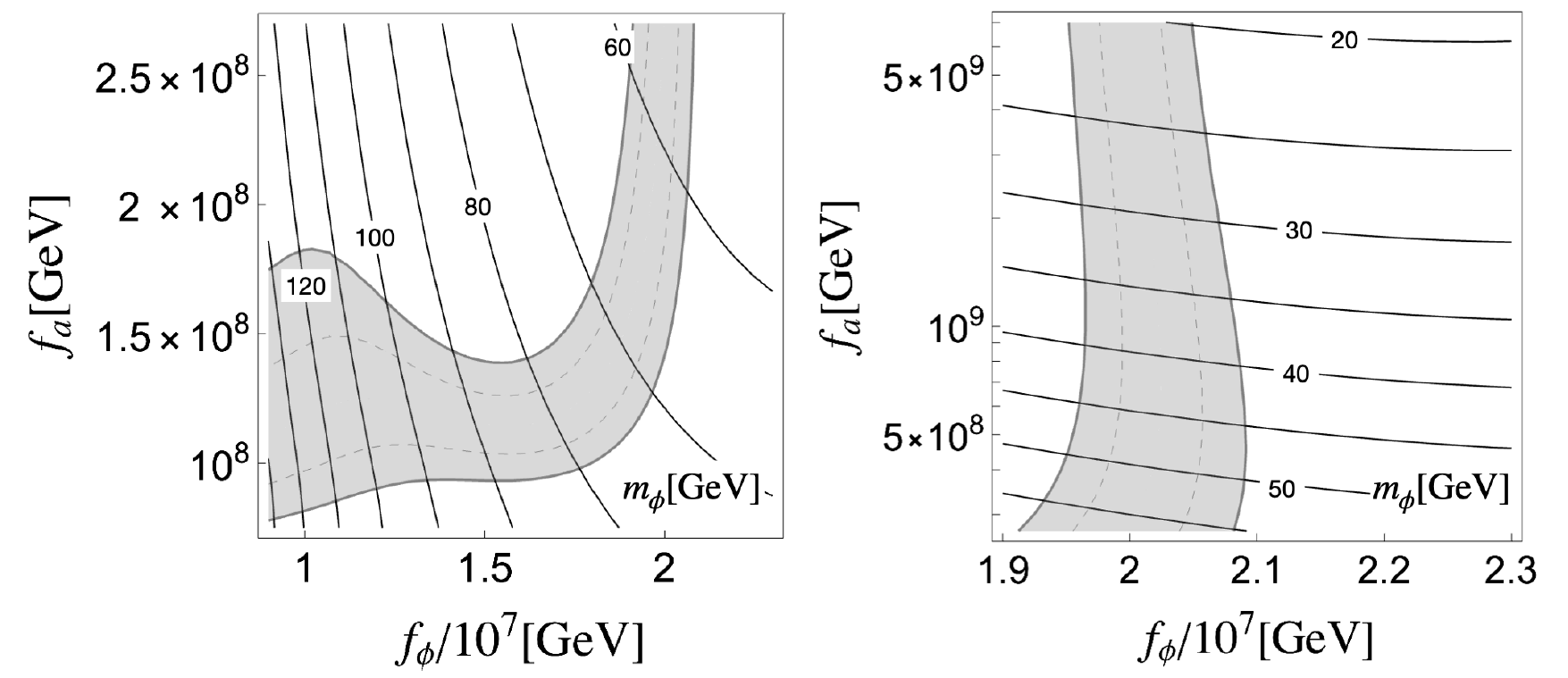}
\end{center}
\caption{The contour plots of $m_\f$ in GeV with respect to $f_\phi$ and $f_a$ incorporating the observational constraints on $n_s$.}
\label{fig:lambdacontour} 
\end{figure}

In this section we discuss a possibility of detecting the ALP. The mass of $\phi$ in the vacuum, $m_\phi$, is estimated by taking the second derivative of the potential with respect to $\f$. For $n=\rm even$, we  have,
\beq
\label{massphi}
m_\f = \frac{\sqrt{2} \L^2}{f_\f},
\eeq
at the potential minimum (\ref{phimin}) and (\ref{amin}).
Since the parameter $\L$ is uniquely determined for given $f_\f$ and $f_a$, we can draw the contour of $m_\phi$ as a function of $f_\phi$ and $f_a$ in Fig.~\ref{fig:lambdacontour}. Thus, the range of ALP mass and decay constants can be narrowed by considering the CMB bound to the QCD axion hybrid inflation.

To see the future prospect of ALP searches at accelerators, let us estimate the production of $\f$ from {gluon-gluon} fusion in a collision of  hadrons, $h_1 \AND h_2$. Its cross section can be estimated as (see Ref.~\cite{Kelly:2020dda})
\beq
\sigma_{h_1h_2\to \f}\sim \frac{\a_s^2 m_\phi^2}{256 \pi f_\f^2 s} I
\eeq
where 
\beq
I=\int^{1}_{m_\f^2/s} \frac{1}{x} f_g^{h_1}[x]f_g^{h_2} [m_\phi^2/(xs)]
\eeq
is the integral of the parton distribution functions of the gluon $g$, and $s$ is the Mandelstam variable. Note that $f_g^{h_i}{[x]}$ 
can be much larger than 1 when $x$ is less than unity, and thus $I$ can be naturally larger than 1. 
The total cross-section for the $h_1 + h_2$ scattering is estimated to be $\rm mb$, and we obtain the production rate of the ALP as
\beq
R\sim \frac{\sigma_{h_1h_2\to \phi}}{\rm mb}\sim 10^{-20} I
 \(\frac{2\times 10^7\GEV}{f_\f}\)^2\frac{\(30\GEV\)^2}{s} \(\frac{m_\f}{25\GEV}\)^2.\eeq
This is a very small production rate, but it may be within the reach of fixed target experiments in the future; with ${\cal O}(10^{22-23})$ protons on target,  one expects $\O$(100-1000) events for the ALP production.  In this case, we may need a beam energy of more than TeV, which is possible with the current technique.

The ALP produced in this way travels a long way. The signal is typically a displaced vertex with two jets, especially for the relatively light ALP.  
One can estimate the decay length of the ALP with energy $E_{\rm ALP}$ to be 
\beq
l_{\rm decay}\sim {10\,\rm m} \left(\frac{E_{\rm ALP}}{1\TEV}\right) \(\frac{f_\f}{2\times 10^7\GEV}\)^2 \(\frac{25\GEV}{m_\f}\)^3,
\eeq 
which is a scale to place a detector with a shielding material behind the collision point. 
So we conclude that in principle the ALP can be probed, but we may need a proper design of a fixed target experiment. 
We also note that the parameter range (mass and decay constant) can be slightly shifted for different values of $n_{\rm mix}$.

\section{Reheating and $N_{\rm eff}$}
\lac{reh}
In this section we discuss the reheating in detail. Towards the end of the inflation, the inflationary trajectory becomes along the ALP direction, and the QCD axion becomes lighter than the Hubble parameter. Thus, the ALP $\phi$ dominates the post-inflationary universe.

After the onset of the oscillation, the $\f$ condensate can be approximated as non-relativistic matter with a mass $m_\f\approx \sqrt{2}\L^2/f_\f$.\footnote{
We set $n=2$ for simplicity. The reheating proceeds in a similar way for other even integers, but will change if $n$ is an odd integer. This case will be mentioned in the last section. 
} The ALP decays into a pair of gluons via the interaction \eq{Lag} at a rate of
\beq
\Gamma_{\rm dec} \approx \frac{ \alpha_{st}^2}{{32}\pi^3} \frac{m_\phi^3}{f_\phi^2}.
\eeq
The gluons immediately form a thermal plasma. The temperature soon becomes higher than $T_{\rm QCD}$, and one can neglect the 
QCD potential.  When the temperature is around the mass of $\f$, the back-reaction from the plasma is important due to the scattering between $\f$ (or gluons) and the ambient plasma.  First, the gluons get the mass of $m^2_{\rm th}\sim \alpha_{st} T^2$ from forward scattering.  Therefore, the decay rate is suppressed by the phase space factor of $\sqrt{1-\left({2m_{\rm th}}/{m_\phi}\right)^2} $. When $m_\f \lesssim 2m_{\rm th}$, the perturbative decays are kinematically forbidden, i.e., the decays are thermally blocked. However, the $\f$ condensate evaporates due the  scattering between $\phi$ and gluons, and the evaporation rate is given by~\cite{Yokoyama:2005dv,Anisimov:2008dz,Drewes:2010pf,Mukaida:2012qn,Drewes:2013iaa,Mukaida:2012bz, Moroi:2014mqa}:
\beq
\laq{sphaleron}
\Gamma^{\rm pert}_{\rm dis} \sim  \frac{\a_{st}^2 T^3}{32\pi^2 f_a^2}  \frac{m_\phi^2}{g_s^4 T^2}.
\eeq
Here the ALP mass $m_\f^2$ appears due to the shift symmetric interaction at the perturbative level. In other words, the ALP coupling to gluons can be rewritten as a derivative couplings of $\f$, which leads to the scattering amplitude proportional to the energy of $\f$. 
On the other hand, if the temperature\footnote{More precisely, the magnetic screening scale  $g_{st}^2 T$.} is higher than $m_\f$,
the QCD sphaleron can also contribute to the evaporation~\cite{McLerran:1990de, Moore:2010jd}
:
\beq
\Gamma^{\rm sph}_{\rm dis} \sim  (N_c\a_{st})^5\frac{T^3}{f_\phi^2}.
\eeq
Here $N_c=3$ is the color factor. This formula is justified in the limit that the oscillating time scale is much faster than the relaxation scale, which will be explained in Appendix.~\ref{app:1}. 
Note that $m_\f^2$ does not appear in this case, because the shift symmetry is broken by the non-perturbative effect of the sphaleron.

\begin{figure}[!t]
\begin{center}  
     \includegraphics[width=113mm]{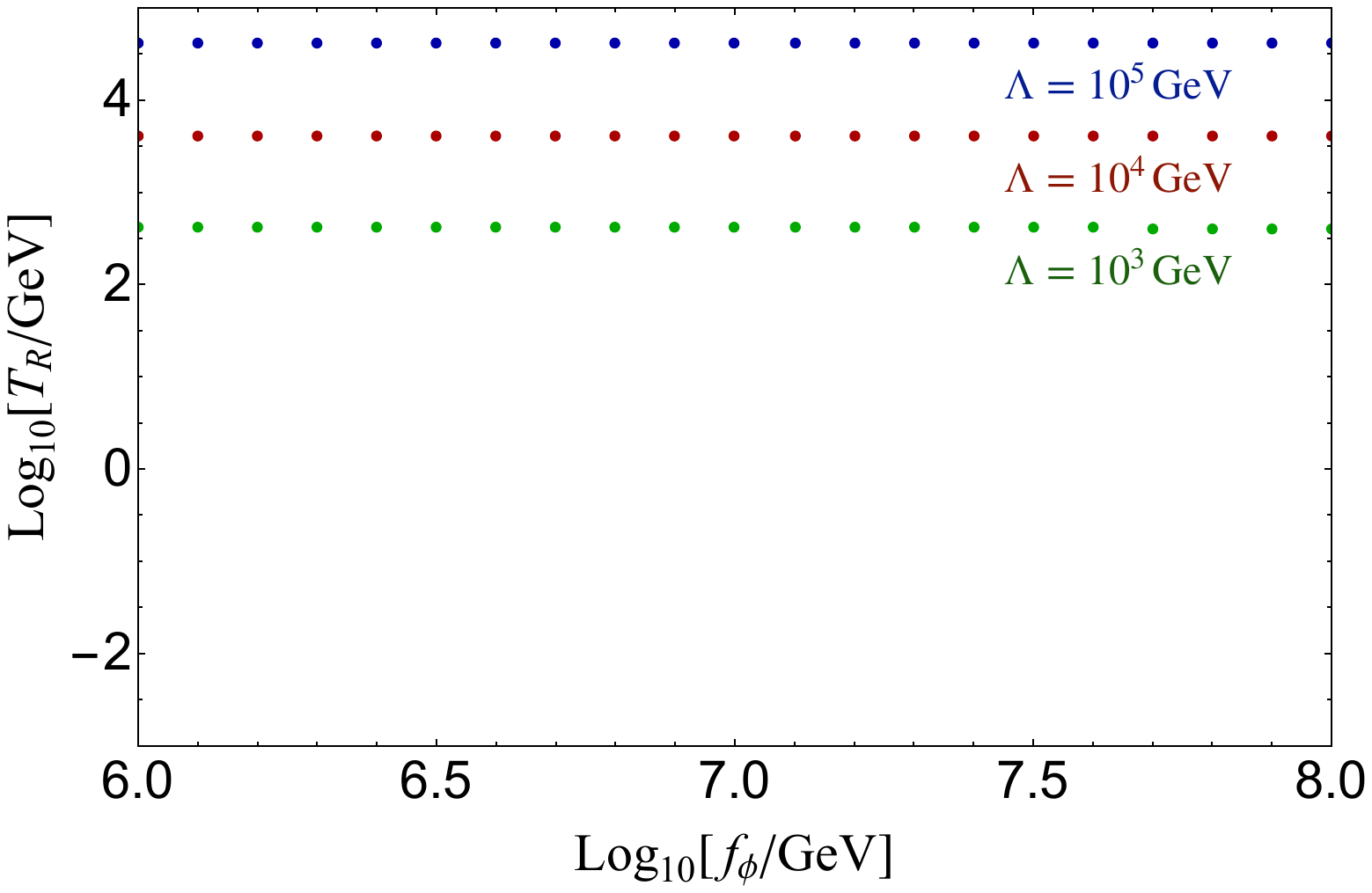}
      \end{center}
\caption{$T_R$ as a function of $f_\f$ for $\L=10^5,10^4,10^3\GEV$ from top to bottom.} \label{fig:5} 
\end{figure}

\begin{figure}[!t]
\begin{center}  
     \includegraphics[width=115mm]{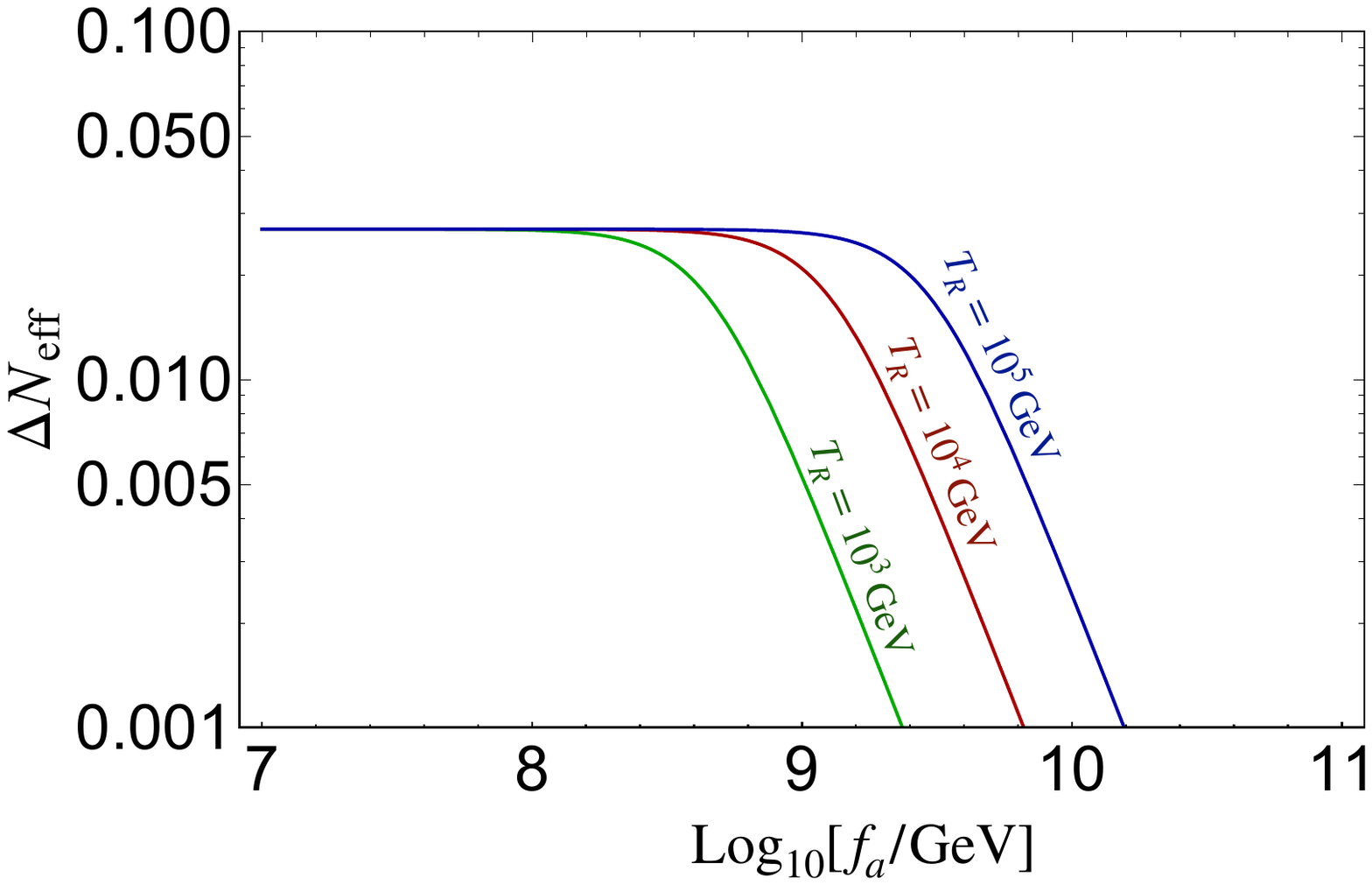}
      \end{center}
\caption{$\Delta N_{\rm eff}$ as a function of $f_a$ for $T_R=10^5,10^4,10^3\GEV$
from top to bottom. }\label{fig:6} 
\end{figure}
In order to estimate the reheating temperature, we solve the Boltzmann equations:
\begin{align}
\left\{\begin{array}{ll}
	\displaystyle{\dot{\rho}_\phi+3H\rho_\phi=-\Gamma_{\rm tot}\rho_\phi} \\
	&\\
	\displaystyle{\dot{\rho}_r+4H\rho_r=\Gamma_{\rm tot}\rho_\phi}\label{evolution}
	\end{array}
	\right.,
\end{align}
where  $\rho_\phi$ and $\rho_r$ denote the energy density of the $\f$ condensate and that of radiation, respectively.  
We take the total decay rate to be the sum of the perturbative decay, dissipation, and sphaleron rates,
\begin{align}
\G_{\rm tot}=
{ \G_{\rm dec}+\G_{\rm dis}^{\rm pert}+\G_{\rm dis}^{\rm sph}}.
\end{align}
 The thermal blocking effect is taken into account using a step function. In the numerical calculation,  we evaluate the reheating temperature at $\r_\phi=\r_\g$. We show the reheating temperature $T_R$ in Fig.~\ref{fig:5} as a function of $f_\f$ for $\L=10^5,10^4,10^3\GEV$ from top to bottom. With $f_\f\ll 10^9\GEV$ we find that the reheating completes soon after the inflation due to the sphaleron effect, and the reheating is instantaneous.  For $f_\f\gg 10^9\GEV$, on the other hand, the reheating completes later due to the perturbative decays of the $\f$ condensate, and the reheating temperature is much lower than $\L$. The reheating temperature is typically below the GeV scale, and the color confinement must be taken into account. Thus, for $\L\sim 10^4\GEV$ and $f_\f\lesssim 10^8\GEV$, which is the parameter range of our interest, the reheating is instantaneous and successful, with all the inflationary energy transferred to radiation. In this case, the reheating temperature is given by
\beq
T_R\approx \(\frac{g_\star \pi^2}{30}\)^{-1/4}V_0^{1/4}.
\eeq

Let us discuss the production of the QCD axion $a$ shortly after reheating.  The QCD potential vanishes immediately after inflation.  The potential of $a$ vanishes and it is frozen by the Hubble friction.  However, the non-zero modes can be produced by thermal scattering.  Such QCD axion particles contribute to the dark radiation (or hot DM) of the universe. Using the numerical results of Ref.\,\cite{Salvio:2013iaa},  we estimate the extra relativistic degrees of freedom, $\D N_{\rm eff}$, by varying $f_a$ for $T_R\sim 10^4\GEV$, which  is shown in Fig.\,\ref{fig:6}.  The $\D N_{\rm eff}$ can be as large as $0.01-0.027$ for $f_a\lesssim 10^{9-10}\GEV$. This can be tested  by future CMB and BAO observations~\cite{Kogut:2011xw, Abazajian:2016yjj, Baumann:2017lmt}.

Let us now focus on the cosmological evolution of this system at the cosmic temperature $T\ll T_R$. Since the reheating is instantaneous, $\f$ settles into the vacuum soon after inflation.  However, $\f$ particles are also thermally produced by the backreaction during evaporation. 
These $\phi$ particles decay perturbatively into gluons.
For $n=2$ (or any other even integer), $\f$ has a mass of 
\beq
\laq{mass}
m_\f\sim 10 \GEV \(\frac{\L}{10\TEV}\)^2 \(\frac{10^7\GEV}{f_\f}\), 
\eeq
 in the vacuum (see Eq.~(\ref{massphi})). The perturbative decay temperature, $T_d\equiv \({\frac{\pi^2g_{\star}}{90}}\)^{-1/4} \sqrt{\G_{\rm dec} M_{\rm pl}}$, 
is estimated by 
\beq
T_d\sim 20\GEV \(\frac{\L}{10\TEV}\)^3 \(\frac{10^7\GEV}{f_\f}\)^{5/2}.
\eeq
Since our scenario requires $f_\f \sim 10^{7}\GEV, m_\f \gtrsim 10\GEV$, the decay occurs much before the big bang nucleosynthesis with $\L \gtrsim 10\TEV$. Also, the decay of the $\phi$ particles does not induce any significant entropy production, since when $\f$ becomes non-relativistic, $\f$ is still in thermal equilibrium due to the fast decay and inverse decay.

\section{QCD axion dark matter}
\lac{DM}

\subsection{Abundance of QCD axion}
Let us estimate the QCD axion abundance in our scenario. In contrast to the usual scenario, the initial misalignment angle is fixed by the inflaton dynamics, since the QCD axion is part of the inflaton and drives the inflation at an early stage. 
The QCD axion begins to oscillate around the potential minimum given by Eq.~(\ref{amin}), when the temperature drops to around the QCD scale and the QCD potential is generated again. 
The coherent oscillation contributes to the {DM energy density}.

The abundance of the QCD axion generated by the misalignment mechanism~\cite{Preskill:1982cy,Abbott:1982af,Dine:1982ah} is given by~\cite{Bae:2008ue, Visinelli:2009zm,Ballesteros:2016xej}
\beq
\laq{ab}
\Omega_a h^2 
\,\simeq\, 
0.0092 F(\h_i )\h_i^{2}
\left(\frac{f_a}{10^{11}\,{\rm GeV}}\right)^{1.17}, 
\eeq
where $\theta_i$ is the initial misalignment angle defined by
$$\theta_i\equiv \frac{|a_i-a_{\rm min}|}{f_a},$$
and  $a_i$ is the initial field value much after the inflation. The function $F(\theta_i)$ is given by
\beq
F(\theta_i)=\left[\log{\(\frac{e}{1-\frac{\theta_i^2}{\pi^2}}\)}\right]^{1.17},
\eeq
which takes into account the anharmonic effect.
So we need to know $a_i$ for given parameters to determine the abundance.
 In the usual case, the observed DM abundance $\Omega_{\rm DM}h^2= 0.120\pm 0.001$~\cite{Aghanim:2018eyx} can be explained with
 $f_a \approx 10^{12}$\,GeV for the initial angle of order unity.
 In our case at hand, we will see below that the anharmonic effect significantly enhances the axion abundance.

{To estimate the position of QCD axion after inflation, we first divide the inflaton dynamics into the periods during inflation and after inflation when ALP oscillates, and then further divide the former into single-field and waterfall periods, and estimate the position of QCD axion at each period.  In the waterfall period, depending on the value of $f_a$, inflation may end immediately or it may continue for a while, and the time when the CMB scale leaves the horizon may differ depending on the parameters. Below we estimate the contribution to $a_i$ for these three periods. We find that the abundance of axion explains the observed DM abundance in the prolonged waterfall regime.}

First let us consider the single-field regime with $f_a \lesssim 10^8$\,GeV. In this case,
the inflaton was almost the QCD axion,
$A_L\approx \frac{\sqrt{1+r_{\rm mix}^2}}{r_{\rm mix}} a\approx a$.
When $A_L \sim A^{\rm cutoff}_L$, the waterfall {period} begins. In the waterfall {period}, the inflaton is almost the ALP, and the QCD axion becomes lighter than the Hubble parameter and does not evolve much with time afterwards. Therefore, we expect that the position of the QCD axion at the onset of oscillations, $a_i$,  is determined by the position where the transition from the single-field {period} to the waterfall {period} occurs.

Next, we assume $f_a\gtrsim 10^{8}\GEV$, and evaluate the evolution of the QCD axion in three different {period}s.
In the single-field {period}, the heavy mass eigenstate is stabilized at the minimum, $A_H\simeq 0$, and $a$ is almost the inflaton, $A_L\approx \frac{\sqrt{1+r_{\rm mix}^2}}{r_{\rm mix}} a\approx a$,  since $r_{\rm mix} \gg 1$.
This relation holds until $A_L \sim A^{\rm cutoff}_L$, i.e.,
\beq
\laq{aearly}
\frac{a_{\rm end}}{f_a} \approx \frac{A^{\rm cutoff}_L}{f_a} \sim 10^{-10}  \times (n^2-1)^{-1/2} \(\frac{10\TEV}{\L}\)^2,
\eeq
where $a_{\rm end}$ denotes the field value of the QCD axion
at the end of  the single-field {period}. Afterward, the system enters the waterfall {period}.

Next, let us follow the evolution of $a$ in the waterfall {period} and the ALP oscillation period.
We can approximate the Hubble parameter to be  constant $H\simeq H_{\rm inf}$ during this period. The mass of the QCD axion is well approximated by the vacuum mass, and it satisfies
\beq m_a=\frac{\sqrt{\chi_0}}{f_a}\ll H_{\rm inf}
\eeq 
because $\L\sim 10^4\GEV.$ 
We can neglect the $\ddot a$ term in the equation of motion and use the slow-roll equation to follow the evolution of the QCD axion, 
\beq
\dot{a} \simeq \frac{\chi_0}{3f_a H_{\rm inf}} \sin{\left(n_{\rm mix}\frac{\f}{f_\f} {-} \frac{a}{f_a} \right)}.
\eeq
Since it is the ALP $\phi$ that mainly moves and $a/f_a$ stays much smaller than unity, we can neglect $a/f_a$ in the parenthesis. Therefore, by integrating this equation of motion, we obtain
\beq
\laq{inta}
\D a\simeq \frac{{\chi_0}}{3f_a {H_{\rm inf}}}  \int dt \sin{\left(n_{\rm mix}\frac{\f}{f_\phi}\right)}.
\eeq

{With respect to the evolution of $\phi$, it can be divided into the waterfall period in which $\phi$ is slow-rolling and the ALP oscillation period in which $\phi$ oscillates around the potential minimum.}
During the slow-roll period of $\phi$, we can approximate the integrand as $\sin{(n_{\rm mix}\frac{\f}{f_\phi})}\simeq n_{\rm mix} \frac{\f}{f_\f}$ since $\phi/f_\phi \ll 1$ in this period.
By changing the integral variable from $t$ to $\phi$ and using the slow-roll equation for $\phi$, we obtain
\beq
\D_{ \text{sr}} a \simeq  {-} \frac{n_{\rm mix}\chi_0}{f_a } \int d \f\, \frac{1}{ {V_\phi'}}\frac{\f }{f_\f} \simeq  \left. \frac{n_{\rm mix}\chi_0}{4f_a }\frac{1}{f_\f \l \f }\right|_{A_L=A_L^{\rm cutoff}}\simeq {\frac{3}{r_{\rm mix}^2}A_{L}^{\rm cutoff}}.
\eeq
Here we have integrated  $\phi$ from the transition point $A_L=A_L^{\rm cutoff}$ until the end of inflation, using the approximation that the integral is dominated by the contribution of the starting point in the second equality. Thus, the field excursion during this period is much smaller than $a_{\rm end}/f_a$ in \Eq{aearly}.

Next, let us estimate the contribution during the ALP oscillation period. In this case, the integrand of the r.h.s of \Eq{inta}  rapidly oscillates around zero. {Since the oscillations in late times cancel out, the main contribution will be during the first few oscillations, when the oscillation amplitude damps significantly  due to Hubble friction.}
We can set an upper bound of the net contribution by taking a $1/4$ oscillation time $\sim 1/m_\f$,
\beq
|\D_{\text{osc}} a|\lesssim 
\frac{\chi_0}{f_a H_{\rm inf} m_\phi},
\eeq
or equivalently,
\beq
\frac{|\D_{\text{osc}} a|}{f_a}\lesssim 10^{-12} \(\frac{10\TEV}{\L}\)^4 \frac{f_\f}{10^7\GEV}\(\frac{10^{9}\GEV}{f_a}\)^2.
\eeq
{This contribution is smaller than $a_{\rm end}/f_a$ for the parameters of our interest. }
In fact, within a few oscillations, the thermalization takes place as we have discussed before. Soon the $\chi(T)$ becomes vanishingly small due to the finite-temperature effects, and thus $a$ is frozen due to the Hubble friction, and its motion can be neglected afterward.

In summary, the initial field value of the QCD axion at the onset of oscillations can be estimated as
\beq 
\label{ai}
a_i =  a_{\rm end}+\Delta_{\text{sr}} a+\Delta_{\text{osc}} a\simeq a_{\rm end}
\eeq 
for the parameters of our interest.
Interestingly, $\frac{a_i}{f_a} \sim n_{\rm mix}^2\frac{\sqrt{\chi_0}}{\L ^2}$ is related to the ratio of the QCD scale to the inflation scale.  
Thus,  the initial misalignment angle of the QCD axion is given by
\beq
\theta_{ i}\simeq
\left\{
\begin{array}{cc}
 \displaystyle{\pi-\frac{{ a_{\rm end}}}{f_a} }  &  ~~{\rm for~~}n_{\rm mix}~ = {\rm ~ odd} \\
 \displaystyle{ -\frac{{ a_{\rm end}}}{f_a}} &  ~~{\rm for~~}n_{\rm mix}~ = {\rm ~ even} 
\end{array}
\right.
 \eeq
The oscillation amplitude is close to $\pi$ when $n_{\rm mix}$ is odd, while it is very suppressed when $n_{\rm mix}$ is even. We focus on the case of odd $n_{\rm mix}$ in the following.

\begin{figure}[!t]
\begin{center}  
     \includegraphics[width=105mm]{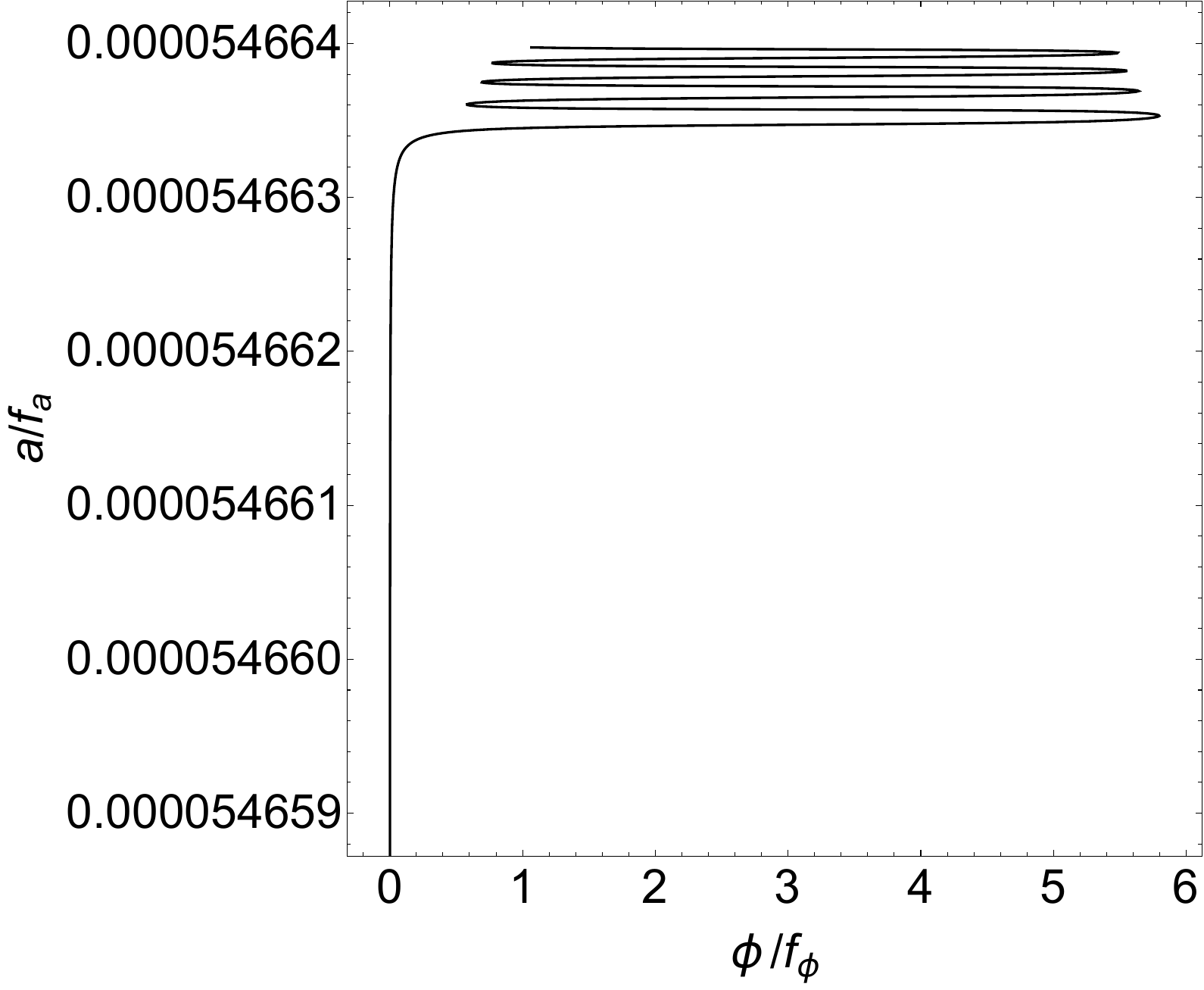}
      \end{center}
\caption{The evolution of $\f$ and $a$  after the inflation for 
 $f_\f=10^{15}\GEV, f_a=10^{17}\GEV, \L=10^{15}\GEV,n=2,n_{\rm mix}=1, \chi_0=\(10^{13}\GEV\)^4$. The choice of these parameters
 is for reducing the numerical cost.
 }\label{fig:DM} 
\end{figure}

To verify that the above estimate is correct, we solved the equation of motion numerically using the full potential until $\f$ oscillates several times. To reduce the numerical cost, we used the following parameters: $f_\f=10^{15}\GEV, f_a=10^{17}\GEV, \L=10^{15}\GEV,n=2,n_{\rm mix}=1$, and $\chi_0=\(10^{13}\GEV\)^4$,
which are different from the values we are interested in.\footnote{We have numerically checked that $a_{\rm end}+\D_{\rm sr} a$ is consistent with the analytical estimate for the parameters of interest.} However, this is sufficient to check the validity of our analytical estimate. The result is shown in Fig.~\ref{fig:DM}.
This parameter set gives $a_{i}/f_a\sim 3\times 10^{-5}$ according to (\ref{ai}), while we get $a_i/f_a\sim 5\times 10^{-5}$ from the numerical result.  We also note that in the few oscillations $a/f_a$ increases by $\O(10^{-10})$, which is consistent with our analytical estimate, $|\D_{\text{osc}} a|/f_a \lesssim 10^{-9}$.

For $n_{\rm mix}=1$ (or any other odd integer), the axion is set around the potential top. 
The QCD axion abundance is estimated as
\beq
\Omega_a h^2 \sim
0.1 \(\frac{F(\h_i )}{50}\)
\left(\frac{f_a}{5\times {10^{9}}\,{\rm GeV}}\right)^{1.17}.
\eeq
Thus the QCD axion can be the dominant DM with $f_a=\O(10^{9})\GEV$, thanks to the anharmonic effect.
We note that the $\O(1)$ uncertainty  in the analytic estimate of {$a_i/f_a$} does not affect this prediction  much because of the logarithmic dependence in \Eq{ab}.
Thus, in order not to overproduce the QCD axion DM one needs 
$
f_a\lesssim {\cal O}(10^{9})\GEV 
$
for odd $n_{\rm mix}$.

This suggested parameter range, $f_a\lesssim {\cal O}(10^{9})\GEV $ is quite interesting from the experimental point of view.  Regardless of whether the QCD axion is the dominant DM,  it can be searched for in the Baby IAXO, IAXO, and IAXO$+$ experiments~\cite{Irastorza:2011gs, Armengaud:2014gea, Armengaud:2019uso} if $f_a < 10^9\GEV$. If $f_a = {\cal O}(10^9)\GEV$ to explain the DM abundance,   it can also be probed in the TOORAD~\cite{Marsh:2018dlj} and BREAD experiments~\cite{BREAD:2021tpx}.   Also the hot axion DM can be probed in the BAO and CMB experiments, as discussed in the previous section. Another important test of the scenario is the heavy $\f$ search, which may require adapting the optimized experimental setup for a beam-dump experiment. The discovery of these phenomena together would strongly support our scenario.   

If $n_{\rm mix}$ is an even integer, on the other hand,  the QCD axion cannot be the dominant DM component due to a too-small misalignment angle. An upper limit on $f_a$ cannot be obtained from the argument of the QCD axion abundance.

\subsection{Axion isocurvature perturbation}
Usually, QCD axion DM has isocurvature perturbations if the PQ symmetry is broken and the QCD axion exists during inflation. 
In particular, when the oscillation starts near the top of the potential, the isocurvature perturbations are known to be enhanced due to the anharmonic effect~\cite{Lyth:1991ub,Kobayashi:2013nva}.  In our scenario, although the anharmonic effect is significant, the isocurvature perturbations are strongly suppressed because the inflation scale is very low. {It is also suppressed at low wavenumbers since the heavy mass eigenstate is so heavy that it can be integrated out, and then the inflaton dynamics are well described by single-field inflation.}

The isocurvature perturbation is defined as follows:
\beq
\mathcal{S} \equiv \frac{\d \Omega_a}{\Omega_{\rm DM}} - \frac{3}{4} \frac{\d \rho_{\gamma}}{\bar{\rho}_{\gamma}}
\simeq \left.\frac{\d \Omega_a}{\Omega_{\rm DM}}\right|_{\delta \rho_\gamma \simeq 0},
\eeq
where $\Omega_{\rm DM}$ denotes the abundance of all DM, $\d \Omega_a$ is the fluctuation of axion density, and in the second equality, we approximate $\delta \rho_{\gamma} \simeq 0$ to be negligible on the uniform density slicing {during the radiation dominated era.} $\d \Omega_a$ is linked to the QCD axion fluctuation $\d a$ as
\beq
\laq{QCDaxionfluc}
\left.\d \Omega_a\right|_{\delta \rho_\gamma \simeq 0} = \frac{\partial{\Omega_a}}{\partial \theta_i} \d \theta_i = \frac{\partial{\Omega_a}}{\partial \theta_i} \frac{\d a}{f_a}.
\eeq
Note that this formula holds regardless of whether the QCD axion is the dominant component DM or not.

\begin{figure}[!t]
\begin{center}  
     \includegraphics[width=0.965\textwidth]{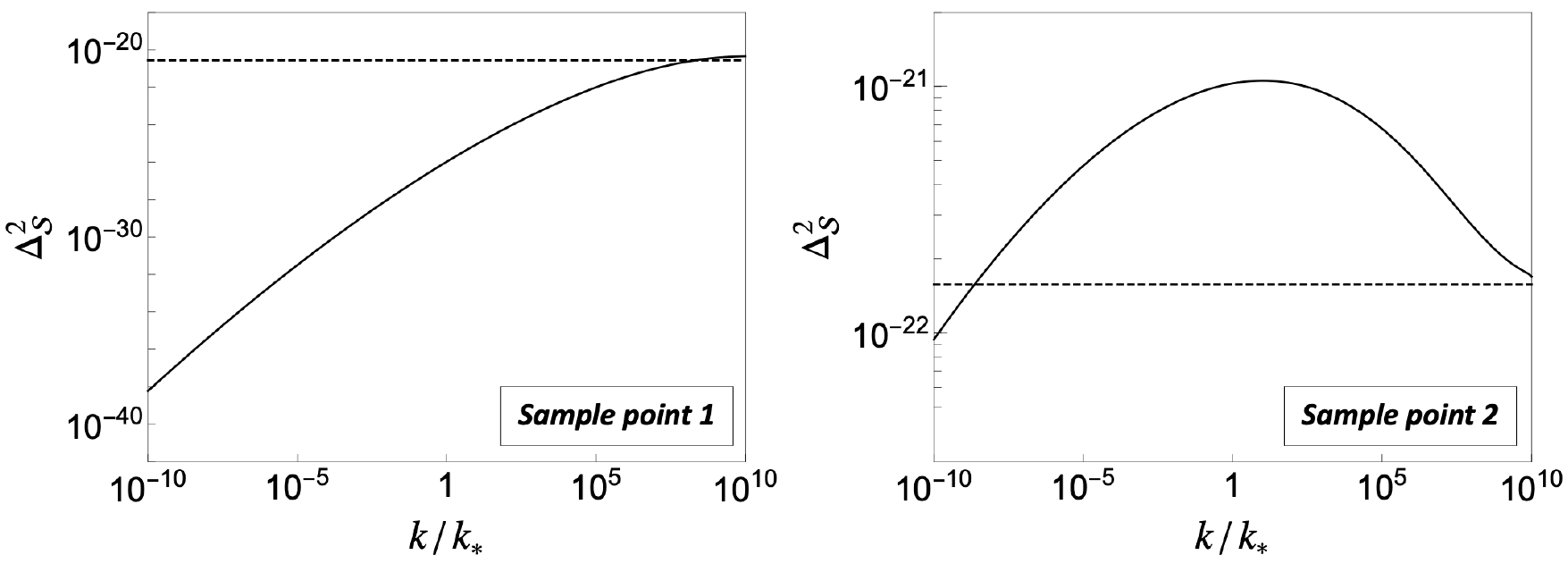}
\end{center}
\caption{The reduced power spectra of isocurvature perturbation $\Delta^2_{\mathcal{S}}(k)$ as a function of $k/k_*$. {The dashed lines represent the analytical estimate Eq.(\ref{anaiso}) at small scales.}}
\label{isops} 
\end{figure}

We performed numerical calculations for the power spectrum of the isocurvature perturbation, $\Delta^2_{\cal S}(k)$, which is shown in Fig.~{\ref{isops}} for the sample point 1 and 2 in the left and right panels, respectively\footnote{
{
In the numerical calculations, we use the longitudinal gauge and estimate $\delta a$ around the end of inflation in the uniform density slicing by $\d a - a'/\phi' \d \phi$,
{which is justified {when} $|a'|\ll |\f'|.$}
}

}. 
At the CMB scale they are
\beq
\Delta^2_{\mathcal{S}}(k_*) \simeq
\left\{
\begin{aligned}
    4.9 \times 10^{-27} \:\:\: \rm for\:sample\:point\:1 \\
    1.7 \times 10^{-21} \:\:\: \rm for\:sample\:point\:2
\end{aligned}
\right. .
\eeq
We can see from the left panel (sample point 1) that the low wavenumber modes are strongly suppressed because they exit the horizon in the single-field period. The high wavenumber modes become larger because the mode exits the horizon when two mass eigenstates are both lighter than the Hubble parameter. In the right panel (sample point 2), this happens around the horizon exit of the CMB scale. So we can have a relatively large isocurvature perturbation, but still too small to be observed.  To have a consistency check, we can analytically estimate the isocurvature perturbation in the large wavenumber limit.  In this limit, the isocurvature mode exits the horizon when the path is almost straight along the ALP, and we can use the usual estimation.  We can estimate the typical scale of $\delta a$ after the trajectory bends,
\beq
\d a \simeq \frac{H_{\rm inf}}{2 \pi}.
\eeq
The power spectrum of isocurvature perturbation at this small scale is calculated as
\beq
\label{anaiso}
\D^{2}_{\mathcal{S}}(k\gg k_*) \simeq 1.6 \times 10^{-22} \left(\frac{F}{50}\right)^{0.29} \left(\frac{10^{-11}}{|\pi - \theta_i|}\right)^2 \left(\frac{H_{\rm inf}}{0.1 \, \rm eV}\right)^2 \left(\frac{f_a}{5 \times 10^9 \rm GeV}\right)^{0.34}.
\eeq
This is shown by the horizontal dashed line, which agrees well with the numerical result.
{Since the isocurvature is very small, the correlated isocurvature perturbation, which should be significant at sample point 2, is also suppressed.  In summary, our scenario is not constrained by isocurvature perturbations, although the QCD axion plays an important role in the inflaton dynamics, including the generation of the curvature perturbations.}

\section{Discussion and conclusions}

\paragraph{Quality of the PQ symmetry and slow-roll condition}

The quality problem of PQ symmetry implies that any explicit PQ breaking terms other than QCD must be extremely suppressed. For example, there may be an explicit PQ breaking term like
\beq
V_{\rm  PQ B}
(a) = \Lambda_{\rm UV}^4 \cos{\left(c_{\rm UV} \frac{a}{f_a} + \theta_{\rm CPV}\right)},
\eeq
where $\L_{\rm UV}$, $\theta_{\rm CPV}$, $c_{\rm UV}$ are the parameters determined by the UV dynamics. 
We expect {that the most natural value of $\theta_{\rm CPV}$ is $\O(1)$.}
Around the origin where the strong CP phase vanishes, we can expand the PQ breaking term as
\beq
\laq{PQB}
V_{\rm  PQ B}\sim \Lambda_{\rm UV}^4 \frac{a}{f_a},
\eeq
where we assume $c_{\rm UV} = {\cal O}(1)$.
Such a PQ breaking term must be suppressed to satisfy
the tight constraint of the neutron electric dipole moment~\cite{Baker:2006ts,Afach:2015sja, Pospelov:2005pr, Dragos:2019oxn},
\beq
\L_{\rm UV}^4\lesssim 10^{-10} \chi_0.
\eeq
A priori there is no reason to suppress the PQ breaking term to such a high degree, and this is the quality problem of the PQ symmetry.

Interestingly, the quality of the PQ symmetry is closely related to the inflationary dynamics, especially the slow-roll condition, in the QCD axion hybrid inflation. 
Indeed,  the linear term affects the slow-roll dynamics and it must  satisfy \cite{Daido:2017wwb,Daido:2017tbr,Takahashi:2019qmh}
\beq
\frac{\L_{\rm UV}^4}{\L^4} \lesssim \O\(\frac{f_a^3}{M_{\rm pl}^3}\),
\eeq
so as not to spoil the hilltop inflation.
This results in
\beq
\frac{\L_{\rm UV}^4}{\chi_0}\lesssim 2\times 10^{-11} \(\frac{f_a}{10^8\GEV}\)^3 \(\frac{\L}{10^4\GEV}\)^4.
\eeq

As we have seen, the decay constant is bounded above as $f_a\lesssim 10^{9} \GEV$, because of the
axion DM bound for odd $n_{\rm mix}$. Then,
the required high quality of the PQ symmetry is at least partially explained by the requirement of successful inflation and the axion DM bound.
For this argument to work, we need to assume that the dominant PQ breaking term can be well approximated as \Eq{PQB}, and that  there is no cancellation between multiple PQ breaking terms. This argument suggests that there should be small but nonzero contributions to
the nucleon EDM.
The induced proton EDM can be searched for in a future proton storage ring experiment~\cite{Anastassopoulos:2015ura,Omarov:2020kws}. 

In the presence of small extra PQ breaking terms, 
the prediction of the curvature power spectrum 
as well as the axion abundance may be slightly modified (see e.g. Ref.~\cite{Nakagawa:2020eeg}). 
In this scenario, the EDM experiment can be linked to the CMB observation and the axion DM in this scenario, which is an interesting future direction.

\paragraph{Case of odd $n$}
 So far we have focused on the even $n$ case, especially $n=2$. When $n$ is odd, the cosmic history remains almost intact during and shortly after inflation, but is changed  at a late time. 
\Eq{mass} can be seen as the effective mass when the amplitude of the $\f$ oscillation is large (see e.g. Refs.~\cite{Daido:2017wwb, Daido:2017tbr}). 
Thus the perturbative decays happen,  followed by the dissipation and sphaleron effects. The reheating should be successful. 
However, the curvature of $V_\f$ vanishes at the minimum.
This means that $a$ is almost massless at the potential minimum, but $\f$ obtains its mass  from the non-perturbative QCD effects.  
Then, $f_\f$ and $m_\f$ have the same relation for the QCD axion. 
In this case, $f_\f\lesssim 10^7\GEV$ is too small to be consistent with astrophysical bounds. 
If we introduce relative phase and height between the two cosine terms in \eq{DIV}, we can avoid this argument. 
Our analysis in this paper then shows that the predicted parameter range can be slightly changed in the presence of the QCD axion by coupling the ALP inflaton to the gluon.
This may provide a new parameter region in the ALP miracle scenario~\cite{Daido:2017wwb, Daido:2017tbr}.

\paragraph{Conclusions}
In this paper we have studied a hybrid inflation driven by the QCD axion $a$.  The key ingredient of our scenario is the mixing between $a$ and another axion (ALP), $\f$, which also couples to the Chern-Simons term of the gluon.  In the vacuum, $a$ gets its mass via the non-perturbative QCD effects, while $\f$ gets its mass via its own potential, which is flat over a limited region away from the vacuum. Around the flat region, a mixing of the two axions becomes relevant, where $\f$ gets most of its mass from QCD and becomes heavier than $a$ and the Hubble parameter. Thus, $\f$ can be effectively integrated out, at least in the early stages of inflation. This regime is called the single-field regime, where
the QCD axion $a$ drives the slow-roll inflation in the effective theory.  In the parameter region satisfying the limits of stellar cooling, the inflation is terminated by a waterfall of $\f$, which can improve the predicted spectral index in a better fit to the CMB observations.   Through the detailed analysis of the two axion fields, we have identified the viable parameter space and shown that the QCD axion can be not only the inflaton but also DM when $f_a = {\cal O}(10^{9})\GEV$.  Interestingly, the abundance of the QCD axion is a function of the ratio of the inflation scale to the QCD axion mass.  This scenario can be probed in the axion direct search experiments such as IAXO, and in future CMB and BAO experiments. In some cases, the {ALP} $\phi$ may be probed by future accelerator experiments. We have also argued that the slow-roll condition partially explains the required high quality of the Peccei-Quinn symmetry.

\section*{Acknowledgments}
We thank Andreas~Ringwald for the useful discussions. W.Y. thanks DESY theory group for hospitality {during the visit} in the fall of 2019. 
This work is supported by JSPS KAKENHI Grant Numbers 17H02878 (F.T.), 20H01894 (F.T.), and 20H05851 (F.T. and W.Y.),  21K20364 (W.Y.),  22K14029 (W.Y.), and 22H01215 (W.Y.), and Graduate Program on Physics for the Universe of Tohoku University (Y.N.). This article is based upon work from COST Action COSMIC WISPers CA21106,  supported by COST (European Cooperation in Science and Technology).

\appendix 

\section{Preliminaries for numerical calculations}
\label{app:nc}
Here we summarize the equations of motion for multiple scalar fields and their fluctuations,  and the definition of the curvature perturbations, the spectral index, and its running.

First, for a general discussion, we introduce canonically normalized {real scalar} fields, $\phi_i$, where $i=1,2,\cdots i_{\rm max}$ is the index of the scalar field. Here $i_{\rm max}$ is the maximum number of fields. 
We decompose $\phi_i$ into the background field, $\bar \f_i$, and the fluctuation, $\d\f_i$, around it:
\beq 
\phi_i(N)= \bar{\phi}_i(N)+\d \phi_i(N).
\eeq
where we consider $\phi_i$ as a function of the e-fold $N$.
The time evolution of the background field is determined by the equations of motion, 
\beq
\frac{\bar \f_i''}{3 - \e_{\phi}} + {\bar\f'}_i+M_{\rm pl}^2\frac{\partial}{\partial \bar\f_i} 
\pqty{\log{V(\bar \f_i)}} = 0, \laq{backgroundfield}
\eeq
where $V(\phi_i)$ denotes the total inflaton potential, and the prime symbol denotes differentiation with respect to the number of e-folds $N$, and the parameter $\e_{\phi}$ is given by
\beq
\laq{slowroll_1e}
\e_{\phi} = \sum_j \frac{\overline{\f}_j'^2}{2 M^2_{\rm pl}},
\eeq
{which becomes equal to the slow-roll parameter $\varepsilon$ in the slow-roll approximation.}

We also need the time evolution of the field fluctuations $\delta \phi_i$ to study the power spectrum of the curvature perturbations. The detailed calculation procedure is described as follows. Using the perturbed equations of motion and the Bardeen equation {in the longitudinal gauge} \cite{Ringeval:2007am}, we can express the equations for their mode functions $\d \f_{i, k}$ as follows:
\begin{align}
\laq{pEoM}
\d \f_{i, k}'' + (&3 - \e_{\phi}) \d \f_{i, k}' + \frac{1}{H^2} \sum_j \pqty{\d \f_{j, k}  \pdvo{\overline{\f}_j}} \pdvt{V}{\overline{\f}_i}  + \frac{k^2}{a^2 H^2} \d \f_{i, k} = 4 \P_k' \overline{\f}_i' - 2\P_k \frac{1}{H^2} \pdvt{V}{\overline{\f}_i},  \\
\laq{Beq1}
&\P_k = \frac{1}{2 M^2_{\rm pl} \(\e_{\phi} - \frac{k^2}{a^2 H^2}\)} \sum_j \sqty{\overline{\f}_j' \d \f_{j, k}' + \pqty{3 \overline{\f}_j' + \frac{1}{H^2}\pdvt{V}{\overline{\f}_j}} \d \f_{j, k}}, \\
\laq{Beq2}
&\P_k' = \frac{1}{2 M^2_{\rm pl}} \sum_j \overline{\f}_j' \d \f_{j, k} - \P_k,
\end{align}
where 
$\P_k$ denotes the mode function of the Bardeen variable.
Note that only two of the three equations Eqs.\eq{pEoM}, \eq{Beq1}, and \eq{Beq2} are independent. In the numerical calculations shown in the main text, we have numerically solved the first two equations.

The Friedmann equation can be written as 
\beq
\laq{Friedmann}
H^2 = \frac{V}{M^2_{\rm pl}(3 - \e_{\phi})}.
\eeq
The mode function of the curvature perturbation is given by
\beq
\laq{curvature_p}
\mathcal{R}_k = - \P_k - \sum_j \frac{\overline{\f}_j' \d \f_{j, k}}{2 M^2_{\rm pl} \e_{\phi}},
\eeq
and we define the power spectrum $P_{\cal R}$ and the reduced power spectrum $\Delta^2_{\mathcal{R}}(k)$  by 
\begin{align}
  \laq{reducedpower}
\vev{\mathcal{R}_k \mathcal{R}^*_{k'}} &= (2 \pi)^3 \d^{(3)}(\vec{k}-\vec{k}') P_{\mathcal{R}}(k), \\
\Delta^2_{\mathcal{R}}(k) &\equiv \frac{k^3}{2 \pi^2} P_{\mathcal{R}}(k)  
\end{align}
where 
$\vev{\cdots}$ means the ensemble average of the fluctuations. {The reduced power spectrum of the isocurvature perturbation, $\Delta_{\cal S}^2$, is defined in a similar way.}

{In the QCD axion hybrid inflation, the power spectrum
should be evaluated when it becomes constant in time after the trajectory bends to the ALP direction. This is because, in the presence of multiple scalar fields, there are isocurvature perturbations that could evolve with time even at superhorizon scales. 

Finally, the spectral index $n_s$, its running $\a_s$, and running of the running $\b_s$ can be extracted by expanding the power spectrum as~\cite{Akrami:2018odb}:
\beq
    \laq{PR}
    \log{\Delta^2_{\mathcal{R}}(k)} = \log{\Delta^2_{\mathcal{R}}(k_*)} + (n_s -1) \log\(\frac{k}{k_*}\) + \frac{1}{2} \a_s \(\log\(\frac{k}{k_*}\)\)^2 + \frac{1}{6} \b_s \(\log\(\frac{k}{k_*}\)\)^3.
\eeq
By expanding the numerically computed power spectrum with such parameters and comparing it to the measured CMB power spectrum, we can narrow down the viable parameter ranges.

\section{QCD sphaleron rate}
\label{app:1}

In the literature, the sphaleron contribution to the friction of a scalar field is often used in a different form. Here we discuss the difference from our setup, where the oscillation time scale $m_\phi$, of the scalar $\phi$ is much faster than the diffusion time scale of the Chern-Simons number, $\Gamma_{\rm sph}\sim N_c^5 \alpha_3^5 T$, averaged over the plasma particle number.

The equation of motion of $\f$ gets altered by the friction in the form with $\SU(N)$ sphaleron with a massless charged fermion is~\cite{Berghaus:2020ekh}
\beq \ddot{\phi}\supset -\frac{\Gamma_{\rm sph}}{2 T }\(\frac{\dot\phi}{f_\phi} -C_R n_{\rm CS}/T^2\). \eeq

Strictly speaking, the Chern-Simons number density, $n_{\rm CS}$ may evolve, following ~\cite{Berghaus:2020ekh}
\beq \dot{n}_{\rm CS}\propto \frac{\Gamma_{\rm sph}}{T }\(\frac{\dot\phi}{f_\phi} -C_R n_{\rm CS}/T^2\). \eeq
The r.h.s has the same factor as the  friction term. 
Thus the sphaleron contribution cancels if the axion rolls are slow enough. This is because then the quasi-equilibrium of $\dot{n}_{\rm CS}=0$   is always reached, and the friction 

However, this is not the case for our case. The oscillation time scale of $\phi$, $m_\phi=\Lambda^2/f_\phi$, is typically of $10^{-3}\Lambda$, which is much faster than the sphaleron time scale $(N_c\alpha_3)^5 T\lesssim 10^{-5}\Lambda$.  Thus the Chern-Simons number density is always zero in the average of the diffusion time scale. 
This is the reason why we can use the form of the sphaleron rate \Eq{sphaleron} in estimating the efficiency of reheating~\cite{Takahashi:2021tff}.\footnote{WY would like to thank Kohei Kamada and Alexandros Papageorgiou for discussions on the sphaleron rate during the Workshop on Very Light DM 2023.}

\section{Solving initial condition problem}
\label{app:2}
Here we present a mechanism for solving the initial condition problem using a temperature-dependent potential for the ALP. This is essentially the same mechanism proposed in the original ALP miracle scenario~\cite{Daido:2017wwb}, but we also comment on the relation to our QCD axion hybrid inflation scenario. There are two types of tuning in small-scale inflation models. The first is the tuning of the potential form. This can be explained by UV completions, as we mentioned, or by the anthropic argument in the string landscape or axiverse.  The other is the initial condition for inflation to take place, i.e. within a Hubble patch we need a  nearly spatially homogeneous value of the inflation field for the slow-roll inflation to take place. {The latter can be explained if another inflation precedes it.}
The amount of fine-tuning associated with the initial condition may also be compensated by the exponentially large volume if the inflation lasts very long. However, the long or eternal inflation is sometimes argued to be problematic due to the appearance of infinities of volume (see e.g. Refs.~\cite{Guth:2000ka,Guth:2007ng,Linde:2015edk}). 

Here we propose a simple mechanism to initially place the inflaton near the top of the potential. We assume that not only the QCD potential but also one of the cosine terms of the inflaton depends on the temperature. For this purpose, we introduce a dark sector with a hidden non-Abelian gauge group. The ALP couples to it with the coupling  
\beq
{\cal L}\supset -\frac{\a_{H}}{8\pi f_\f} \phi G_H \tl{G}_H,
\eeq
where $\a_H$  is the hidden gauge coupling, and $G_H$ is the field strength. We assume that the gauge coupling becomes strong on the energy scale around $\Lambda$, so that the first cosine term of \eq{DIV} is formed. This implies that the cosine term depends on the temperature of the hidden sector, $T_H$,
 \begin{align}
\label{eq:DIV2} 
V_{\f}(\phi) = \Lambda^4 \(\k_1(T_{H})\cos\(\frac{\phi}{f_\f} + \Theta \)- \frac{\kappa }{n^2}\cos\(n\frac{\f}{f_\f }\)\)+{\rm 
const.}
\end{align}
{where $\kappa_1$ denotes the temperature-dependent coefficient, and $\Theta$ is much smaller than unity for successful inflation.}
If the phase transition is of second order, $\k_1$ is a continuous function of the temperature, and it can be parametrized by 
\beq
\k_1(T)= \min{[(T/T_c)^p, 1]},
\eeq
 where $p(<0)$ is a number of order unity that depends on the details of the matter content and mass. 
$T_c$ is the critical temperature that satisfies $T_c\sim \L.$. The temperature dependence  is similar to the QCD axion potential.

 Let us assume that before inflation the universe is full of plasma of the hidden sector with $T_H\gg T_c$.
 In this case $\kappa_1$ is vanishingly small, and thus the potential of $\phi$ is mainly composed of the second cosine function.  For a while we do not consider the mixing of $\f$ with $a$, which we can easily justify because the temperature is naturally higher than the QCD scale.  Then,  $\phi$ is driven to the minimum with $\phi=0$.  As the universe cools, the first cosine term becomes important, and the ALP is set to the top of the hill. The potential can be expanded around $\phi = 0$ as
 \begin{align}
\label{eq:expH} 
V_{\f}(\phi) = {\rm const.} + \L^4 \k_1(T_H) \Theta  \frac{\f}{f_\f}+ \frac{m^2(T_H)}{2} \f^2 + \cdots,
\end{align}
where $m^2(T_H)=(\k-\k_1(T_{H})) \frac{\L^4}{f_\f^2} \sim \max [p(T_H/T_c-1) , 0]\frac{\L^4}{f_\f^2}$ for $\k\simeq 1$.   Then the mass goes from $\O(\L^2/f_\f)$ to almost zero within $\O(1)$ Hubble time.  Since the former is much larger than $H$ and the latter much smaller than $H$, $\f$ would follow the minimum until $m^2(T_H)\sim H^2$. This corresponds to the field value of 
\beq
\f\sim \frac{\H \L^4 }{ f_\f H^2} \sim \H \frac{M_{\rm pl}^2}{f_\f}.
\eeq
In the last equality, we used $H\sim \L^2/M_{\rm pl}$. 
On the other hand, the slow-roll regime of quartic hilltop inflation is 
\beq
|\f|\lesssim \frac{f_\f^2}{M_{\rm pl}}.
\eeq 
So we get the condition that the ALP is set near the potential maximum where the slow-roll inflation takes place,
\beq
\Theta \lesssim \(\frac{f_\f}{M_{\rm pl}}\)^3.
\eeq 
This condition is checked numerically by solving the equation of motion with a time-varying potential.  In the context of ALP inflation~\cite{Daido:2017wwb, Daido:2017tbr, Takahashi:2019qmh}, this corresponds to the regime of hilltop ALP inflation. 

A potential problem with this mechanism for setting an initial condition for the QCD axion hybrid inflation is that it does not constrain the field value of the QCD axion.  We can assume that the nearly homogeneous and flat universe is large enough due to the previous inflation, which can be a high-scale inflation.  Suppose that the previous inflaton reheats the universe to a very high temperature.  Although the $\phi$ field is set to $\phi\simeq 0$ by the above mechanism,  the QCD axion has position-dependent random values with a flat distribution over the whole universe. 

Noticing that the non-vanishing field value $a \ne 0$ effectively induces a linear term of $\phi$, we get an additional contribution from $\frac{n_{\rm mix}\chi_0}{f_\phi} \sin[a/f_a]$ in the equation of motion for $\phi$ after the universe cools down below the QCD scale. Here the term $n_{\rm mix}\phi/f_\phi\sim 0$ in the argument is neglected. Only at the field points where $a/f_a$ is small enough, we can have a second inflation (i.e. the QCD axion hybrid inflation), because otherwise it would violate the slow-roll condition of $\phi$. In the region where $a/f_a$ is small enough, it settles down to the minimum (the attractor solution we discussed) on a time scale of 
\beq \Delta N \sim \frac{H_{\rm inf}^2}{m_a^2} \sim 8000 \(\frac{H_{\rm inf}}{0.1 \EV}\)^2 \(\frac{f_a}{5\times 10^{9}\GEV}\)^2\eeq 
which is not too long. Then the QCD axion hybrid inflation begins.

\bibliographystyle{apsrev4-1}
\bibliography{reference}

\end{document}